\documentclass[12pt,a4paper,titlepage]{article}
\pdfoutput=1                          
\usepackage[utf8x]{inputenc}
\usepackage[english]{babel}

\usepackage{ucs}
\usepackage{amsmath}
\usepackage{amsfonts}
\usepackage{amssymb}
\usepackage{eucal}
\usepackage{mathrsfs}          
\usepackage[sans]{dsfont}     
\usepackage{slashed}
\usepackage[Symbol]{upgreek}  
\usepackage{array}            
\usepackage{cite}               
\usepackage[normalem]{ulem}             
\usepackage{booktabs}      
\usepackage{colortbl}         
\usepackage{fancyhdr}
\usepackage{multirow}
\usepackage{feynmf}           
\usepackage
{graphicx}

\usepackage{color}
\definecolor{grigio}{cmyk}{0,0,0,0.1}
\definecolor{rosa}{cmyk}{0,0.1,0.1,0.02}
\definecolor{rosino}{cmyk}{0,0.05,0.05,0.02}
\definecolor{rosas}{cmyk}{0,0.3,0.25,0.05}
\definecolor{celeste}{cmyk}{0.1,0,0,0.02}
\definecolor{giallino}{cmyk}{0,0,0.1,0.02}
\definecolor{rosso}{cmyk}{0,1,1,0.4}
\definecolor{rossos}{cmyk}{0,1,1,0.55}
\definecolor{rossoc}{cmyk}{0,1,1,0.2}
\definecolor{blu}{cmyk}{1,1,0,0.3}
\definecolor{blus}{cmyk}{1,1,0,0.5}
\definecolor{bluc}{cmyk}{1,1,0,0.1}
\definecolor{blucc}{cmyk}{0.7,0.5,0,0}
\definecolor{viola}{cmyk}{0,1,0,0.6}
\definecolor{viola2}{cmyk}{0,1,0.2,0.6}
\definecolor{verde}{cmyk}{0.92,0,0.59,0.25}
\definecolor{verdec}{cmyk}{0.92,0,0.59,0.15}
\definecolor{verdes}{cmyk}{0.92,0,0.59,0.4}
\definecolor{verdino}{cmyk}{0.12,0,0.09,0.02}
\definecolor{giallo}{cmyk}{0,0,1,0}
\definecolor{gialloverde}{cmyk}{0.44,0,0.74,0}
\definecolor{Titolo}{rgb}{0.752941176,0.576470588,0.992156863}
\definecolor{altro}{rgb}{0.094117647,0.650980392,0.643137255}
\definecolor{Peanuts}{rgb}{0.2, 0.4, 0.6}
\definecolor{Pean1}{rgb}{0.6, 0.8, 0.4}
\definecolor{BHO}{rgb}{0.2, 0.8, 1}
\definecolor{Daria}{rgb}{0, 0.9412, 0}
\definecolor{UniPi}{rgb}{0.2549, 0.4627, 0.6275}
\definecolor{UniPidue}{rgb}{0.3216, 0.5804, 0.7882}
\definecolor{rossoCP3}{cmyk}{0,.88,.77,.40}
\definecolor{verdeCP3}{rgb}{0.09765625, 0.57421875, 0.1015625}
\definecolor{bluCP3}{rgb}{0, 0.23, 0.67}
\definecolor{bluSaclay}{rgb}{0, 0.22, 0.70}

\usepackage{cancel}
\usepackage[normalem]{ulem}

\newcommand{\spazio}{\bigskip}
\newcommand{\eg}{e.g.~}
\newcommand{\ie}{i.e.~}

\newcommand{\Op}{{\cal O}}		
\newcommand{\Lag}{\mathscr{L}}	
\newcommand{\Mel}{\CMcal{M}}	

\newcommand{\virg}[1]{`#1'}

\newcommand{\uno}{\mathds{1}}
\font\bb=bbmss10 scaled 1200
\newcommand{\unop}{\mbox{\bb 1}}
\newcommand{\ud}{\text{d}}
\newcommand{\beq}{\begin{equation}}
\newcommand{\eeq}{\end{equation}}

\newcommand{\Eq}[1]{Eq.~\eqref{#1}}
\newcommand{\Fig}[1]{Fig.~\ref{#1}}
\newcommand{\Sec}[1]{Sec.~\ref{#1}}
\newcommand{\App}[1]{Appendix \ref{#1}}
\newcommand{\Ref}[1]{Ref.~\cite{#1}}           

\newcommand{\ER}{E_\text{R}}
\newcommand{\Ed}{E'}
\newcommand{\vesc}{v_\text{esc}}
\newcommand{\vmin}{v_\text{min}}

\newcommand{\NN}[1]{\langle N | #1 | N \rangle}
\newcommand{\NR}{\text{NR}}

\newcommand{\aS}{\alpha_\text{s}}

\long\def\symbolfootnote[#1]#2{\begingroup\def\thefootnote{\fnsymbol{footnote}}\footnote[#1]{#2}\endgroup}

\ifx\pdfoutput\undefined
\usepackage[dvips,bookmarks]{hyperref}    
\else
\usepackage{hyperref}    
\fi
\hypersetup{colorlinks, bookmarksopen, bookmarksnumbered,
citecolor=verdeCP3, linkcolor=bluCP3, pdfstartview=FitH, urlcolor=rossoCP3}

\def\hhref#1{\href{http://arxiv.org/abs/#1}{#1}} 
\def\mhref#1{\href{mailto:#1}{#1}}        

\def\XENON{{\sc Xenon100}}
\def\CDMS{{\sc Cdms-Ge}}
\def\COUPP{{\sc Coupp}}
\def\PICASSO{{\sc Picasso}}
\def\LUX{{\sc Lux}}
\def\SCDMS{{\sc SuperCdms}}

\begin{document}

\begin{titlepage}

\begin{flushright}
\scriptsize
SACLAY--T13/022 \hfill
CP$^3$-Origins-2013-014 DNRF90 \hfill
DIAS-2013-14
\end{flushright}
\color{black}
\vspace{0.3cm}

\begin{center}
{\LARGE{\color{rossoCP3}\bf Tools for model-independent bounds\\ in direct dark matter searches}\\[5mm]
\color{black}
{\Large \bf (updated including LUX 2013 data)\\[-2mm]
(updated including SuperCDMS 2014 data)
}\rule{0pt}{25pt}}
\end{center}

\par \vskip .2in \noindent

\begin{center}
{\sc Marco Cirelli$\, {\color{bluCP3}^{1}}$\,\symbolfootnote[1]{\mhref{marco.cirelli@cea.fr}},
Eugenio Del Nobile$\, {\color{bluCP3}^{2,3}}$\,\symbolfootnote[2]{\mhref{delnobile@physics.ucla.edu}},
Paolo Panci$\, {\color{bluCP3}^2}$\,\symbolfootnote[3]{\mhref{panci@cp3-origins.net}}}
\end{center}

\begin{center}
\par \vskip .1in \noindent
{\it ${\color{bluCP3}^1} \, $\href{http://ipht.cea.fr/en/index.php}{Institut de Physique Th\'eorique}, CNRS, URA 2306 \& CEA/Saclay,\\ F-91191 Gif-sur-Yvette, France}
\\
{\it ${\color{bluCP3}^2} \, $\href{http://cp3-origins.dk}{CP$\,^3$-Origins and DIAS}, University of Southern Denmark,\\
Campusvej 55, DK-5230 Odense M, Denmark}
\\
{\it ${\color{bluCP3}^3} \, $\href{http://www.pa.ucla.edu}{UCLA}, Department of Physics and Astronomy,\\ 
475 Portola Plaza, Los Angeles, CA-90095, USA}
\par \vskip .5in \noindent

\end{center}

\begin{center}
{\large Abstract}
\end{center}

\begin{quote}
We discuss a framework (based on non-relativistic operators) and a self-contained \href{http://www.marcocirelli.net/NRopsDD.html}{set of numerical tools} to derive the bounds from some current direct detection experiments on virtually any arbitrary model of Dark Matter elastically scattering on nuclei.

\end{quote}

\end{titlepage}

\newpage

\tableofcontents

\newpage

\section{Introduction}

Since several decades, cosmological and astrophysical evidences accumulate on the existence of Dark Matter (DM), thanks to its gravitational influence at different scales: at the galactic scale, with the flattening of rotation curves, at the galaxy cluster scale, with weak lensing measurements, and the cosmological scale, with CMB (Cosmic Microwave Background) and LSS (Large Scale Structure) observations. See \eg \cite{Bertone:2004pz} for a review of all these aspects. 
On the other hand, an explicit (non gravitational) manifestation of DM is yet to be identified. As well known, there are three main approaches: direct detection searches aim at detecting a feeble kick to an atomic nucleus in extremely low background environments; indirect detection techniques aim at unveiling possible excess cosmic rays produced by the annihilations or decay of DM particles in the Galaxy or beyond; collider searches aim at identifying signatures of the production of the DM particles at high energy particle accelerators, most notably the Large Hadron Collider (LHC). 

\medskip

The observables which are relevant in each one of these approaches, of course, all pertain to the interactions of the unknown Dark Matter with the ordinary matter, which is what we measure `in the laboratory'. In direct detection (DD), the important quantity is the cross section for scattering of DM particles off a nucleon, in a deeply non-relativistic regime (since the speed of the incoming DM particle is of the order of $v \sim 10^{-3} c$, with $c$ the speed of light) characterized by a momentum transfer of a few (tens to hundreds of) MeV. As a consequence, as stressed recently by~\cite{Fan:2010gt,Fitzpatrick:2012ix,Fitzpatrick:2012ib}, the most useful framework for DD analyses is in terms of a basis of {\em non-relativistic operators} (given in Eq.~(\ref{NRoperators}) below), which parameterize most efficiently the actual interactions among DM and ordinary matter in terms of a very limited number of non-relativistic degrees of freedom.
This approach turns out to be very generic and powerful, given that, in spite of the many possible DM candidates one can envision (complex or real scalar, Dirac or Majorana fermion, vector), and of the different interactions they can have with matter, the scattering process can always be expressed in terms of a very limited number of non-relativistic degrees of freedom. As a matter of fact, one can write a basis of non-relativistic (NR) operators; the matrix element of {\em any} process of elastic scattering between the DM and the nucleon, regardless of the high energy interaction responsible for the scattering, can then be expressed as a combination of these NR operators. Moreover, most of the field operators usually encountered in both renormalizable and effective theories reduce to only a subset of the NR operators, given in Eq.~(\ref{NRoperators}) below.

\medskip

In order to make contact with the actual experimental observables (for instance: the number of events in a detector like \XENON), however, one needs to perform two additional steps. Firstly, one has to compute the nuclear response caused by each one of the NR operators, or a combination of those. This can be done using the formalism clearly spelled out in Fitzpatrick {\it et al.}~in~\cite{Fitzpatrick:2012ix}. Their work provides {\em form factors} $F_{i, j}^{(N, N')}$ which take into account the non-relativistic physics of the DM-nucleus interaction, and encode all the nuclear information as well as the dynamics of the DM-nucleus interaction, for a number of different nuclei of interest in DD experiments. We will heavily rely on their results, as we will discuss in the following.
Secondly, one has to concretize the computed nuclear response to a specific experiment, considering detection efficiency, cuts, acceptances and the energy resolution of the detector. Moreover, one has to make astrophysical assumptions on quantities such as the density and velocity distribution of DM particles in the local halo, and fold them in the computation. This is where we come in: we will present a formalism in terms of {\em integrated form factors} $\tilde{\mathcal{F}}_{i, j}^{(N,N')}$ which allow to compute the constraints from DD experiments taking into account the concrete experimental details and the astrophysical assumptions. This constitutes the main result of our work.

\medskip

Now, Dark Matter being a particle physics problem, any `model' of the interactions of DM with ordinary matter will be expressed in terms of operators involving the degrees of freedom proper of the high-energy theory, \ie quarks and gluons. If one works in the framework of a specific theory comprising a DM candidate (\eg Supersymmetry or Extra Dimensions), one generally knowns the explicit form of such terms, however complicated and parameter-loaded they might be. Conversely, if one is ignorant or agnostic about the underlying theory, one can express the interaction `Dark Matter -- ordinary matter' in terms of {\em effective operators}, which are descriptions of particle interactions at energies lower than the masses of the interaction mediators. 
In both cases, anyway, the high-energy physics will necessarily reduce to NR operators.
Indeed, in the second part of our paper, we make this reduction explicit within the relativistic effective operator approach, reminding how one passes from a set of general high-energy effective operators to the low-energy NR operators. 

\medskip

Summarizing, the goal of our paper is to describe a streamlined method and to provide the needed tools to compute the bounds from the main current DD experiments without having to perform explicitly the steps described above. 
The starting point can be a description in terms of relativistic effective operators (in which case we provide a self-contained guide to obtaining the bounds) or any high energy theory, in which case it is up to the model builder to reduce it to the relevant NR operators on which our tools can be employed. 
The concrete output consists of a set of Test Statistic (TS) functions, one per each featured experiment, and a set of scaling functions (based on the integrated form factors) allowing to translate a benchmark bound into a bound for an arbitrary combination of operators. 
These numerical products are provided on  \href{http://www.marcocirelli.net/NRopsDD.html}{this website: {\tt www.marcocirelli.net/NRopsDD.html}}. It is worth stressing that these results apply mainly to the case of DM-nuclei elastic scattering. However, our method is quite general and in principle similar tools can also be developed for other classes of models (\eg inelastic DM scattering). We encourage therefore independent extensions in this direction.

\bigskip

\begin{table}
\centering
\begin{tabular}{c|c|c|c}
Experiment & Section & Reference & Updated as of \\
\hline
\XENON & \ref{XENON} & \cite{Aprile:2012nq,Aprile:2013doa}  & \multirow{6}{*}{03.2014}  \\
\CDMS & \ref{CDMS} & \cite{Ahmed:2009zw} & \\
\COUPP & \ref{COUPP} & \cite{Behnke:2012ys} & \\
\PICASSO & \ref{PICASSO} & \cite{Archambault:2012pm} & \\
\cline{1-3}
\LUX & Add.~\ref{LUX} & \cite{Akerib:2013tjd} & \\
\cline{1-3}
\SCDMS & Add.~\ref{SuperCDMS} & \cite{Agnese:2014aze} & 
\end{tabular}
\caption{\em \small \label{tableexp} {\bfseries List of the experiments} that we consider, the section where they are discussed and the corresponding references.}
\end{table}

Before moving on, three important remarks concerning the experiments are in order. (i) In this paper we focus exclusively on bounds from DD experiments which have reported null results. Of course there is another set of DD experiments whose results can be interpreted as claims for an allowed DM region. While a method similar to ours can be constructed for the analysis of these experiments, we leave this to future work. (ii) The TS functions that we provide for each experiment reflect the current status of results: they will change whenever new results are released and we will update them accordingly. On the other hand, the integrated form factor (and, a fortiori, the scaling functions) change only if major changes occur in the experimental set-ups, \eg in efficiencies, cuts, and thresholds. We do not foresee frequent updates for those, but we will consider them when necessary. (iii) Among the many very well performing experiments, we choose to focus on six of them, since these currently provide the strongest bounds. Our method, however, obviously can be applied to any set-up and indeed we encourage independent extensions to other experiments. Table~\ref{tableexp} (updated in the arXiv version of this paper) lists the experiments we consider with the corresponding most updated reference.

\medskip

The rest of this paper is organized as follows. In Sec.~\ref{sec:formfactors} we lay down the formalism of the NR operators and illustrate how the experimental observables are expressed in terms of it. 
In Sec.~\ref{sec:constraints} we discuss how one derives constraints based on the experimental results and how one can rescale a general benchmark bound into a bound on a given DM model.
In Sec.~\ref{sec:experiments} we describe in some detail the experiments that we consider, in order to present the specific form that the integrated form factors take for each one of them.
In Sec.~\ref{sec:dictionary} we discuss the reduction from the set of high-energy effective operators to the low-energy NR operators.
In Sec.~\ref{sec:examples} we illustrate all the formalism with some explicit examples.
Finally in Sec.~\ref{sec:conclusions} we conclude.

\section{Phenomenology: from the NR operators formalism to the experimental observables}
\label{sec:formfactors}

In this Section we introduce the formalism of NR operators, following~\cite{Fan:2010gt, Fitzpatrick:2012ix}, and we describe how to compute the experimental observables (essentially the number of events in the energy bins of a certain experiment) in terms of it. More precisely, we will write the differential event rate as a linear function of a manipulation of the {\em  form factors} $F_{i, j}^{(N, N')}$ (provided by~\cite{Fitzpatrick:2012ix}), which take into account the non-relativistic physics of the DM-nucleus interaction, and encode all the nuclear information as well as the dynamics of the DM-nucleus interaction.

\medskip

In a NR description of the elastic scattering of a DM particle $\chi$ with a nucleon $N$, the relevant degrees of freedom are the DM-nucleon relative velocity $\vec{v}$, the exchanged momentum $\vec{q}$, the nucleon spin $\vec{s}_N$ and the DM spin $\vec{s}_\chi$ (if different from zero). The scattering amplitude will then be a rotationally invariant function of these variables; invariance under Galilean boosts is ensured by the fact that these vectors are by themselves invariant under Galileo velocity transformations, and translational symmetry is also respected given the absence of a reference frame/point in space. In this regard, a basis of $16$ rotationally invariant operators can be constructed with $\vec{v}$, $\vec{q}$, $\vec{s}_N$, and $\vec{s}_\chi$ \cite{Dobrescu:2006au}, which include all possible spin configurations. The scattering amplitude can then be written as a linear combination of these operators, with coefficients that may depend on the momenta only through the $q^2$ or $v^2$ scalars ($\vec{q} \cdot \vec{v} =  q^2 / 2 \mu_N$ by energy conservation, with $\mu_N$ the DM-nucleon reduced mass). Before introducing these NR operators, however, let us notice that, instead of $\vec{v}$, the variable $\vec{v}^\perp \equiv \vec{v} - \vec{q} / 2 \mu_N$ is somehow more suitable to write the amplitude. $\vec{v}^\perp$ is Hermitian, in a sense explained \eg in \Ref{Fitzpatrick:2012ix}, while $\vec{v}$ is not, and moreover one has $\vec{v}^\perp \cdot \vec{q} = 0$. Following \Ref{Fitzpatrick:2012ix} we will therefore use, in the description of the NR operators, $\vec{v}^\perp$ instead of $\vec{v}$. The NR operators considered in this work are
\begin{equation}
\label{NRoperators}
\begin{aligned}
\Op^\NR_1 &= \unop \ ,
&
&
\\
\Op^\NR_3 &= i \, \vec{s}_N \cdot (\vec{q} \times \vec{v}^\perp) \ ,
&
\Op^\NR_4 &= \vec{s}_\chi \cdot \vec{s}_N \ ,
\\
\Op^\NR_5 &= i \, \vec{s}_\chi \cdot (\vec{q} \times \vec{v}^\perp) \ ,
&
\Op^\NR_6 &= (\vec{s}_\chi \cdot \vec{q}) (\vec{s}_N \cdot \vec{q}) \ ,
\\
\Op^\NR_7 &= \vec{s}_N \cdot \vec{v}^\perp \ ,
&
\Op^\NR_8 &= \vec{s}_\chi \cdot \vec{v}^\perp \ ,
\\
\Op^\NR_9 &= i \, \vec{s}_\chi \cdot (\vec{s}_N \times \vec{q}) \ ,
&
\Op^\NR_{10} &= i \, \vec{s}_N \cdot \vec{q} \ ,
\\
\Op^\NR_{11} &= i \, \vec{s}_\chi \cdot \vec{q} \ ,
&
\Op^\NR_{12} &= \vec{v}^\perp \cdot (\vec{s}_\chi \times \vec{s}_N) \ ,
\end{aligned}
\end{equation}
where we follow the numbering adopted in \Ref{Fitzpatrick:2012ix, Fitzpatrick:2012ib}. As in \cite{Fitzpatrick:2012ix}, we do not consider the full set of independent operators (for instance, as apparent, we do not consider the operator labeled $\Op^\NR_{2}$ in \cite{Fitzpatrick:2012ix}, nor those above the 12$^{\rm th}$); however, as we will see in \Sec{sec:dictionary}, the operators listed above are enough to describe the NR limit of many of the relativistic operators often encountered in the literature. The form factor for the operator $\Op^\NR_{12}$ was obtained from the authors of \cite{Fitzpatrick:2012ix,Anand:2013yka}.

\bigskip

Given a model for the interaction of DM with the fundamental particles of the SM, we can build the non-relativistic effective Lagrangian describing DM-nucleon interactions as follows. Starting from the fundamental Lagrangian, the matrix element for a scattering process at the nucleon level\footnote{Note that this quantity coincides with what is denoted as a Lagrangian $\mathcal{L}$ in \cite{Fitzpatrick:2012ix, Fitzpatrick:2012ib}, \eg in Eq.~(55) of \cite{Fitzpatrick:2012ix}.} can be expressed as a linear combination of the operators \eqref{NRoperators}:
\beq\label{Leff}
\Mel_N =
\sum_{i = 1}^{12} \mathfrak{c}^N_i(\lambda, m_\chi) \, \Op^\NR_i \ .
\eeq
The coefficients $\mathfrak{c}^N_i$, where $N = p, n$ can be proton or neutron, are real functions of the parameters of the model, such as couplings, mediator masses and mixing angles, (collectively denoted) $\lambda$, the DM mass $m_\chi$ and the nucleon mass $m_N$. 
For example, if the scattering between a fermonic DM $\chi$ and the nucleon $N$ is described by the (high-energy) scalar operator $g_N/\Lambda^2 \ \bar \chi \chi \, \bar{N} N$, the only non-relativistic operator involved is $\Op^\NR_1$, and its coefficient is $\mathfrak{c}^N_1 = 4 \, g_N m_\chi m_N/\Lambda^2$. The general way to determine the coefficients entering the matrix element \eqref{Leff}, starting from high-energy effective operators, will be described explicitly in \Sec{sec:dictionary}.

As anticipated above, the $\mathfrak{c}^N_i$ can in principle also depend on the exchanged momentum squared $q^2$; in this case we factorize the momentum dependence outside of the coefficients and redefine the $\mathfrak{c}^N_i$ as independent from $q$. The most notable cases of $q$ dependence is featured perhaps in long range interactions, where the exchange of a massless mediator is responsible for the interaction between the DM and the nucleons. The differential cross section displays in this case negative powers of $q$, thus enhancing the scattering rate at lower exchanged momenta. Assuming that the massless mediator responsible for the interaction is the Standard Model photon,\footnote{As we shall see in \Sec{sec:dictionary}, gluons behave differently and need separate treatment.} the most relevant cases are a DM with small but nonzero electric charge, electric dipole moment or magnetic dipole moment. As we shall see in more detail in \Sec{sec:dictionary}, these interactions all feature a $1/q^2$-dependence~\cite{Fitzpatrick:2012ib}.
In addition to those in Eq.~(\ref{NRoperators}), we will therefore consider also the following {\it l}ong {\it r}ange operators:

\begin{equation}
\begin{aligned}
\Op^\text{lr}_1 = \frac{1}{q^2} \, \Op^\NR_1 \ , \qquad
&
\Op^\text{lr}_5 = \frac{1}{q^2} \, \Op^\NR_5 \ ,
\\
\Op^\text{lr}_6 = \frac{1}{q^2} \, \Op^\NR_6 \ , \qquad
&
\Op^\text{lr}_{11} = \frac{1}{q^2} \, \Op^\NR_{11} \ .
\end{aligned}
\end{equation}

\medskip

According to Eq.~(55) of \cite{Fitzpatrick:2012ix} we can then write the spin-averaged amplitude squared for scattering off a target nucleus $T$ with mass $m_T$ as
\beq\label{Mel}
\overline{\left| \Mel_T \right|^2} = \frac{m_T^2}{m_N^2} \sum_{i, j = 1}^{12} \sum_{N, N' = p, n} \mathfrak{c}^N_i \mathfrak{c}^{N'}_j F_{i, j}^{(N, N')} \ .
\eeq
The $F_{i, j}^{(N, N')}(v, \ER, T)$ are the form factors provided in the appendices of \cite{Fitzpatrick:2012ix}, and depend critically on the type of scattering nucleus $T$; they are also functions of $m_\chi$, $v$ and the nuclear recoil energy $\ER = q^2 / 2 m_T$.
\\
We can then construct the differential scattering cross section, which reads, in the non-relativistic case,
\beq\label{sigmaT}
\frac{\ud \sigma_T}{\ud \ER} (v, \ER) = \frac{1}{32 \pi} \frac{1}{m_\chi^2 m_T} \frac{1}{v^2} \overline{| \Mel_T |^2} \ .
\eeq
To write the scattering rate we need to take into account the general case in which the detector is composed of different nuclides (these can be different isotopes of the same specie, as well as different kind of nuclei).
We take the numeric abundances of different nuclides used in direct detection searches from Table II of \cite{Feng:2011vu}, and convert them into mass fractions\footnote{$\xi_T = 10^3 N_\text{A} m_T \zeta_T / {\rm kg} \, \bar{A}$, where $N_\text{A} = 6.022 \times 10^{23}$ is Avogadro's number, $\zeta_T$ are the numeric abundances and $\bar{A} \equiv \sum_T \zeta_T A_T$.} $\xi_T$ for each type of target nucleus $T$, with mass number $A_T$, in the detector. The differential rate for DM scattering off a specific target, expressed in cpd (counts per day) per kilogram per keV, is then
\beq\label{RT}
\frac{\ud R_T}{\ud \ER} = \frac{\xi_T}{m_T} \frac{\rho_\odot}{m_\chi} \int_{\vmin(\ER)} \hspace{-.60cm} \ud^3 v \, v \, f_\text{E}(\vec v) \frac{\ud \sigma_T}{\ud \ER} (v, \ER) \ ,
\eeq
where $\rho_\odot \simeq 0.3$ GeV/cm$^3$ is the DM energy density at the Earth's location and $f_\text{E}(\vec v)$ is the DM velocity distribution in the Earth's frame. $\vmin(\ER)$, the minimum velocity with which a DM particle can scatter off a nucleus with a given recoil energy $\ER$, also depends on the target nucleus via the relation $\vmin = \sqrt{m_T \ER / 2 \mu_T^2}$ (for elastic scattering), where $\mu_T$ is the DM-nucleus reduced mass.
\\
Using \Eq{Mel} we get the following expression for the differential rate:
\beq\label{Rate}
\frac{\ud R_T}{\ud \ER} =
X \, \xi_T \sum_{i, j = 1}^{12} \sum_{N, N' = p, n} \mathfrak{c}^N_i(\lambda, m_\chi) \, \mathfrak{c}^{N'}_j(\lambda, m_\chi) \, \mathcal{F}_{i, j}^{(N, N')}(\ER, T) \ ,
\eeq
where we defined
\beq
X \equiv \frac{\rho_\odot}{m_\chi} \frac{1}{32 \pi} \frac{1}{m_\chi^2 m_N^2} \ .
\eeq
and
\beq
\mathcal{F}_{i, j}^{(N, N')}(\ER, T) \equiv \int_{\vmin(\ER)} \hspace{-.50cm} \ud^3 v \, \frac{1}{v} \, f_\text{E}(\vec v) \, F_{i, j}^{(N, N')}(v, \ER, T) \ .
\eeq
In this work we use the customary Maxwell-Boltzmann distribution for the DM velocity. We fix the velocity dispersion to $v_0 = 220$ km/s and we truncate the distribution at $\vesc = 544$ km/s \cite{Smith:2006ym}. We refer to Appendix \ref{sec:velocityntegral} for definitions and details. The important thing to notice here is that, up to first order in the non-relativistic expansion of the cross section, one can encounter only two types of velocity dependence, namely $\ud \sigma_T / \ud \ER \propto v^{-2}$ and $\ud \sigma_T / \ud \ER \propto v^{0}$ (the latter being actually a velocity in-dependence). For these two cases, we use the two integrals $\mathcal I_0(\vmin)$ and $\mathcal I_1(\vmin)$ defined in \Eq{I0} and \eqref{I1}, respectively. Our formalism also allows to treat the case of a linear combination of these two velocity dependences, that one finds \eg in scattering of DM candidates with magnetic dipole moment.

The linearity of $\ud R_T / \ud \ER$ in the form factors is a fundamental ingredient that allows to parametrize a scattering rate in terms of few functions $F_{i, j}^{(N, N')}$. All the operations we will perform on this quantity will preserve such linearity and will enable us to provide a simple recipe to \virg{scale} a bound for a specific DM-nucleus interaction to the appropriate bound for another type of interaction.

\medskip

To actually compare \Eq{Rate} with the experimental rate, we now have to take into account detection efficiency, cuts acceptance and energy resolution of the detector. The set of operations that allows to go from the differential rate \eqref{Rate} to the measured rate is very different from one experiment to the other, and will be discussed in detail later on for each of the experiments considered here (see \Sec{sec:experiments}). For the sake of introducing our results, however, let us sketch now an illustrative procedure.\\
The differential rate must be convolved with the (target-dependent) probability $\epsilon(\Ed)$ $ \mathcal{K}_T(\ER, \Ed)$ that a recoil energy $\ER$ is measured as $\Ed$, taking also into account the quenching. $\epsilon(\Ed)$ is understood to be the detector's efficiency and acceptance, while $\mathcal{K}_T(\ER, \Ed)$ reproduces the effect of the energy resolution in every bin, and it is usually assumed to be a Gaussian distribution (with possibly energy-dependent width). In the end we have the differential rate as a function of the detected energy $\Ed$:
\beq
\label{dRdEd}
\frac{\ud R}{\ud \Ed} = \sum_{T} \epsilon(\Ed) \int_0^\infty \ud \ER \, \mathcal{K}_T(\ER, \Ed) \, \frac{\ud R_T}{\ud \ER} \ .
\eeq
After convolving with all the experimental effects, the recoil rate of \Eq{dRdEd} must be averaged over the energy bin of the detector. For each energy bin $k$ of width $\Delta E_k$, therefore, we define the number of events predicted by the theory to be
\beq
\label{trueNumberOfEvents}
N^\text{th}_k = w_k R_k \equiv w_k \int_{\Delta E_k} \hspace{-.10cm} \ud \Ed \, \frac{\ud R}{\ud \Ed} \ ,
\eeq
with $w_k$ the exposure (expressed in kg$\, \cdot \,$days) and $R_k$ the expected rate in the $k^\text{th}$ energy bin. This quantity can then be directly compared with the detected number of events in the same bin, once the expected number of background events is taken into account.

Collecting all the elements in the previous equations we expand \Eq{trueNumberOfEvents} and write
\beq\label{Nkth}
N^{\rm th}_k = X \sum_{i, j = 1}^{12} \sum_{N, N' = p, n} \mathfrak{c}^N_i(\lambda, m_\chi) \, \mathfrak{c}^{N'}_j(\lambda, m_\chi) \, \tilde{\mathcal{F}}_{i, j}^{(N,N')}(m_\chi, k) \ .
\eeq
Here $\tilde{\mathcal{F}}_{i, j}^{(N,N')}(m_\chi, k)$ is a sort of {\em integrated form factor} that encapsulates all the information related to nuclear physics, astrophysics and the detector dependency of the rate. There is one of these factors for each energy bin $k$ of each experiment under consideration, and for each choice of pair of operators $i,j$ and a pair of nucleons $N,N'$. 
In the approximative determination of the experimental rate sketched above, this would read explicitly
\beq
\tilde{\mathcal{F}}_{i, j}^{(N,N')}(m_\chi, k) = w_k \sum_T \xi_T
\int_{\Delta E_k} \hspace{-.10cm} \ud \Ed \, \epsilon(\Ed) \int_0^\infty \ud \ER \, \mathcal{K}_T(\ER, \Ed) \, \mathcal{F}_{i, j}^{(N, N')}(\ER, T) \ .
\eeq
While this definition has only illustrative purposes, our analysis will take into detailed account the process of deriving the $\tilde{\mathcal{F}}_{i, j}^{(N,N')}$'s for each experiment as shown in \Sec{sec:experiments}.

At this point one can write the total number of events predicted by the theory as
\beq\label{Nth}
N^\text{th} = \sum_k N^\text{th}_k = X \sum_{i, j = 1}^{12} \sum_{N, N' = p, n} \mathfrak{c}^N_i(\lambda, m_\chi) \, \mathfrak{c}^{N'}_j(\lambda, m_\chi) \, \tilde{\mathcal{F}}_{i, j}^{(N,N')}(m_\chi) \ ,
\eeq
where we defined $\tilde{\mathcal{F}}_{i, j}^{(N,N')}(m_\chi) \equiv \sum_k \tilde{\mathcal{F}}_{i, j}^{(N,N')}(m_\chi, k)$. Due to the linearity of the operations performed to get to \Eq{Nth}, the quantity $N^\text{th}$ inherits from \Eq{Rate} the possibility of being expressed as a linear combination of few ingredients, here the \virg{integrated} form factors $\tilde{\mathcal{F}}_{i, j}^{(N,N')}$. 

The meaning of Eq.~\ref{Nth} is that, once one has computed or is provided with the integrated form factor functions $\tilde{\mathcal{F}}_{i, j}^{(N,N')}$, one can straightforwardly derive the expected number of events from possibly {\em any} model of DM interactions, whose particle physics is entirely encapsulated in the $\mathfrak{c}^N_i$ coefficients, and to immediately compare it with the experimental results.
This is the subject of the next Section.

\section{Deriving and rescaling constraints}
\label{sec:constraints}

In the previous Section we have seen how to write down the prediction for the experimental observables (the number of events per bin) in terms of NR operators and the corresponding form factors. 
In this Section we show how to esplicitly use that formalism in order to derive a bound, from experimental data, on the physics parameters $\lambda$ entering in the formalism (see \eg \Eq{Nth}) \ie ultimately on the DM particle physics. 
First we derive a bound for a benchmark model. Then we discuss how to translate that result into a bound for an arbitrary choice of operators, \ie an arbitrary model. 
Before all this, however, it is useful to review the basics of the statistical method employed to derive constraints, specialized to the case at hand.

\medskip

For experiments that do not see anomalies in their counting rate, the bounds on the free parameter(s) of a given DM model arise by comparing the theoretical number of events with the data, taking also into account the predicted background. As already said, in general direct detection experiments can employ more than one detector (module), or be sensitive to nuclear recoils in different energy bins. For each module or energy bin labelled by $k$, we can write the total number of expected events as
\beq
N_k(\lambda, m_\chi) = N^\text{th}_k(\lambda, m_\chi) + N_k^{\rm bkg} \ ,
\eeq
where $N_k^{\rm bkg}$ is the expected number of background events. The predicted number of events from DM $N^\text{th}_k$ depends both on the DM mass $m_\chi$ and on the parameters $\lambda$, including \eg coupling constants and mediator masses.\footnote{From now on we will consider $\lambda$ as representing exclusively those parameters whose value is unknown, and that one wishes to constrain using direct detection searches.}

To compare the theoretical model with the data, we use a standard Likelihood approach. Since for null result experiments the number of events is very low, data are distributed according to a Poisson distribution. Upon assuming a flat prior, the likelihood of obtaining the set of experimentally observed data $\vec N^\text{obs}$ given a certain value of the unknown parameter(s) $\lambda$ is
\beq
\mathcal L(\vec N^\text{obs} \, | \, \lambda) =
\prod_k \frac{ N_k(\lambda, m_\chi)^{N_k^\text{obs}}}{N_k^\text{obs} !} \, e^{- N_k(\lambda, m_\chi)} \ .
\eeq
It is more convenient to use the log of the Likelihood function, that writes
\beq
- 2 \ln \mathcal L(\vec N^\text{obs} \, | \, \lambda) = - 2 \sum_k N_k^\text{obs} \ln N_k + 2 \sum_k \ln \left(N_k^\text{obs} ! \right) + 2 \sum_k N_k \ .
\eeq

Supposing that $\lambda = 0$ corresponds to zero rate, \ie no DM signal, we define the background likelihood $\mathcal L_{\rm bkg} \equiv \mathcal L(\vec N^\text{obs} \, | \, 0)$. To compare the DM model with the background-only hypothesis, we adopt the Likelihood ratio Test Statistic
\beq
{\rm TS}(\lambda, m_\chi) = - 2 \ln \left( \mathcal L(\vec N^\text{obs} \, | \, \lambda) / \mathcal L_{\rm bkg} \right) \ ,
\eeq
which according to the theory has an approximated $\chi^2$ distribution with a number of degrees of freedom equal to the number of free parameters $\lambda$ of the DM model. We can then extract, for any given $m_\chi$, the maximal value of the parameter(s) $\lambda$ allowed by the dataset $\vec N^\text{obs}$. Explicitly one has
\beq\label{Deltachi2Poiss}
{\rm TS}(\lambda, m_\chi) = - 2 \sum_k N^\text{obs}_k \ln \left( \frac{N^\text{th}_k(\lambda,m_\chi) + N_k^{\rm bkg}}{N_k^{\rm bkg}} \right) + 2 \sum_k N^\text{th}_k(\lambda,m_\chi) \ .
\eeq
Bounds on the parameter(s) $\lambda$ at a given confidence level (CL) can be determined by solving ${\rm TS}(\lambda, m_\chi) = \chi^2_{\rm CL}$, where the right-hand side is the $\chi^2$ value corresponding to the desired CL; for instance a 90\% CL bound is obtained by choosing $\chi^2_{90\%} = 2.71$, for one single parameter $\lambda$. A contour plot in the $(m_\chi, \lambda)$ parameter space is the standard way to numerically solve this equation, yielding at the same time a graphical presentation of the result.

\medskip

\begin{figure}[t]
\begin{center}
\includegraphics[width=.5\textwidth]{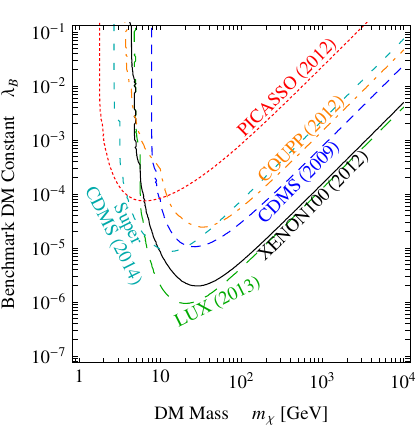}
\end{center}
\caption{\em \small \label{fig:lambdaB}
{\bfseries Our benchmark constraints}, \ie the constraints at 90\% CL on the benchmark DM constant $\lambda_{\rm B}$ entering \Eq{eq:LagrangianBenchmark} as a function of the DM mass from \XENON\ (solid black), \CDMS\ (dashed blue), \COUPP\ (dashed-dotted orange), \PICASSO\ (dotted red), \LUX\ (long-dashed green) and \SCDMS (short-dashed dark cyan).}
\end{figure}

Let us now fix a {\it benchmark model} of DM interaction with the nuclei. $\lambda_\text{B}$ will denote the free parameter of this model, while $\lambda_{\rm B}(m_\chi)$ will denote the bound on it to a given CL, for each value of the DM mass. The total number of expected events for the benchmark model will be denoted as $N^\text{th,B}$. Any model can be used as benchmark, and we choose here the simplest possible one: 
\begin{equation}
\label{eq:LagrangianBenchmark}
\Mel_{p,\text{B}} = \lambda_\text{B} \, \Op^\NR_1 \hspace{2cm} {\rm (benchmark)},
\end{equation}
\ie $\mathfrak{c}^p_1 = \lambda_\text{B}$ while all the other $\mathfrak{c}^N_i = 0$. This results obviously in $N^\text{th,B} = X \, \lambda_\text{B}^2 \, \tilde{\mathcal{F}}_{1,1}^{(p,p)}(m_\chi)$. From this it is possible to solve ${\rm TS}(\lambda_\text{B}, m_\chi) = \chi^2_{\rm CL}$ for a given experiment to a given CL, thus determining the bound $\lambda_\text{B}(m_\chi)$. On \href{http://www.marcocirelli.net/NRopsDD.html}{the website} we provide the user with the functions ${\rm TS}(\lambda_{\rm B}, m_\chi)$ for the four experiments we consider as Mathematica interpolating functions; for a chosen CL, a contour plot in the $(m_\chi, \lambda_\text{B})$ plane draws the bound $\lambda_\text{B}(m_\chi)$ as a function of $m_\chi$, as illustrated in Fig.~\ref{fig:lambdaB}.

\medskip

We can now scale this benchmark bound to a bound on the parameter $\lambda$ of another DM-nuclei interaction, at the same CL. First one can notice that, once the bound $\lambda_{\rm B}(m_\chi)$ for the benchmark model is given, a bound on $\lambda$ (as a function of $m_\chi$) for another interaction is trivially obtained by equating
\beq\label{TS}
{\rm TS}(\lambda, m_\chi) = {\rm TS}(\lambda_{\rm B}(m_\chi), m_\chi) \ ,
\eeq
where the right-hand side is computed within the benchmark model for any value of the DM mass. For an experiment counting the total number of recoil events, the general and unique solution to this equation is $N^\text{th} = N^\text{th,B}$. We can immediately see that this apparently trivial result is actually very powerful: in fact, this translates into
\beq
\label{sol1}
\sum_{i, j = 1}^{12} \sum_{N, N' = p, n} \mathfrak{c}^N_i(\lambda, m_\chi) \, \mathfrak{c}^{N'}_j(\lambda, m_\chi) \, \mathcal{Y}_{i, j}^{(N,N')}(m_\chi) = \lambda_\text{B}(m_\chi)^2 \ ,
\eeq
where the functions
\beq\label{Y}
\mathcal{Y}_{i, j}^{(N,N')}(m_\chi) \equiv \frac{\tilde{\mathcal{F}}_{i, j}^{(N,N')}(m_\chi)}{\tilde{\mathcal{F}}_{1,1}^{(p,p)}(m_\chi)}
\eeq
are provided on \href{http://www.marcocirelli.net/NRopsDD.html}{the website} as Mathematica interpolating functions. With also the ${\rm TS}(\lambda_{\rm B}, m_\chi)$ interpolating functions in hand, a bound on $\lambda$ is promptly obtained by drawing the contour plot of ${\rm TS}(\lambda_{\rm B}(m_\chi), m_\chi)$ at the desired CL, using the expression for $\lambda_{\rm B}(m_\chi)$ in \Eq{sol1}.
The $\mathcal{Y}_{i, j}^{(N,N')}$ functions are plotted for illustration in figures \ref{fig:Yieqj} to \ref{fig:YineqjLR} below. Notice that, due to the interplay of form factors with different signs for different nuclides, the integrated form factors (and thus the $\mathcal{Y}$'s) can be positive or negative depending on the DM mass. In order to accommodate the $\mathcal{Y}$'s in logarithmic plots, we therefore show only their absolute value in the figures.

For experiments measuring a spectrum in the recoil energies, instead of the mere total event number, rigorously \Eq{TS} translates into an equality between the total number of events $N^\text{th} = N^\text{th,B}$ (summed over all bins) only if the predicted spectrum is flat, \ie the expected number of signal events is the same in each bin. However, \Eq{sol1} still provides an excellent approximation to the true bound computed with \Eq{Deltachi2Poiss} for the experimental data sets employed here. We verified this explicitly for the operators presented in \Sec{sec:formfactors} for which the maximum difference in number of events in different bins is expected.

\medskip

Summarizing, on \href{http://www.marcocirelli.net/NRopsDD.html}{the website} we provide:
\begin{itemize}
\item The function ${\rm TS}(\lambda_{\rm B}, m_\chi)$ (as a function of the DM mass $m_\chi$ over the range 1 GeV $\to$ 10 TeV and of $\lambda_{\rm B}$ over the range $10^{-15}$ $\to$ $10^{5}$) that allows the user to compute the bound $\lambda_\text{B}^\text{CL}(m_\chi)$ at the desired confidence level.
\item The functions $\mathcal{Y}_{i, j}^{(N,N')}(m_\chi)$ for each experiment that we consider (\XENON, \CDMS, \COUPP, \PICASSO), for each pair of indices $(i,j)$ and each pair $(N,N')$, as functions of the DM mass $m_\chi$ over the range 1 GeV $\to$ 10 TeV.
\item A sample Mathematica notebook which illustrates the usage of the above files.
\end{itemize}
With these ingredients, \Eq{sol1} allows to set a bound on a parameter $\lambda$ for {\em any} possible interaction type (meaning any possible choice of the $\mathfrak{c}^N_i$). If $\lambda$ consists of a set of several parameters, one can analogously draw constraints on one parameter with the others fixed, on combination of parameters (if factorizable) or even obtain multidimensional bounds.

A frequent case is when $\lambda$ is an overall multiplicative parameter that sets the scale of the cross section (and therefore of the rate). This is, for instance, often the case when $\lambda$ is a coupling constant or a product of coupling constants, or when $\lambda = \Lambda^{-n}$ with $\Lambda$ a scale of new physics or the mass of the interaction mediator (if one process only dominates). This means that we can set $\mathfrak{c}^N_i(\lambda, m_\chi) = \lambda \, \mathfrak{c}^N_i(m_\chi)$ in \Eq{Leff}, so that \Eq{sol1} reads
\beq
\label{soloverall}
\lambda_\text{B}(m_\chi)^2 = \lambda(m_\chi)^2  \sum_{i, j = 1}^{12} \sum_{N, N' = p, n} \mathfrak{c}^N_i(m_\chi) \, \mathfrak{c}^{N'}_j(m_\chi) \, \mathcal{Y}_{i, j}^{(N,N')}(m_\chi) \ .
\eeq
If the analytic form of the benchmark bound $\lambda_\text{B}(m_\chi)$ is known, this equation allows to obtain the analytic bound on $\lambda(m_\chi)$. However, the procedure sketched above for obtaining a bound on $\lambda$ by evaluating the ${\rm TS}(\lambda_{\rm B}(m_\chi), m_\chi)$ with \Eq{sol1} is fully general, and provides fast numerical results if one is able to draw the contour plot of the ${\rm TS}$ at the desired CL.

In Sec.~\ref{sec:examples} we will illustrate how to apply \Eq{sol1} in the context of different models and different choices of the parameter $\lambda$.

\section{Description of the experiments}
\label{sec:experiments}

In this Section we describe in detail the experiments that we consider, in order to present the specific form that the integrated form factors $\tilde{\mathcal{F}}$ take for each one of them. In practice, the following subsections articulate for each experiment the qualitative discussion presented in the second part of Sec.~\ref{sec:formfactors}. The \LUX\ and \SCDMS\ experiments are discussed in the Addenda at page \pageref{LUX} and \pageref{SuperCDMS}, respectively. The uninterested reader can skip this Section and just consider the $\tilde{\mathcal{F}}$ functions (and consequently the $\mathcal{Y}$ functions) as black boxes to be plugged into Eq.~(\ref{sol1}).

\begin{figure}[t]
\parbox[b]{0.76\linewidth}{
\includegraphics[width=0.36\textwidth]{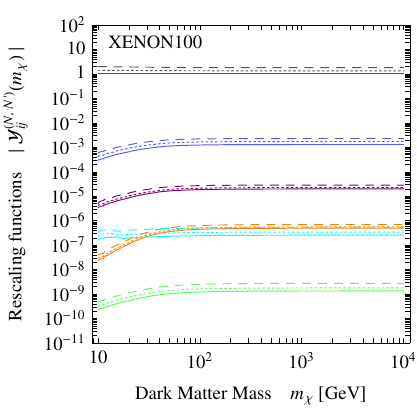} \quad
 	\includegraphics[width=0.36\textwidth]{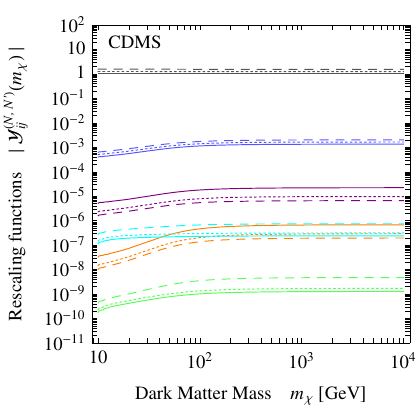}  \\
\includegraphics[width=0.36\textwidth]{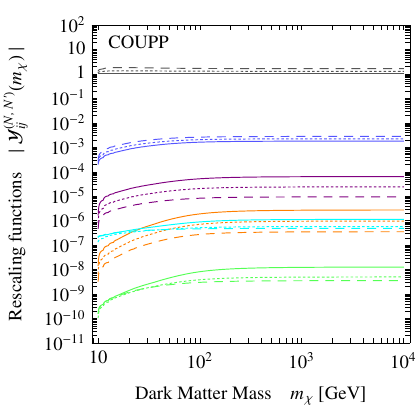} \quad 
 	\includegraphics[width=0.36\textwidth]{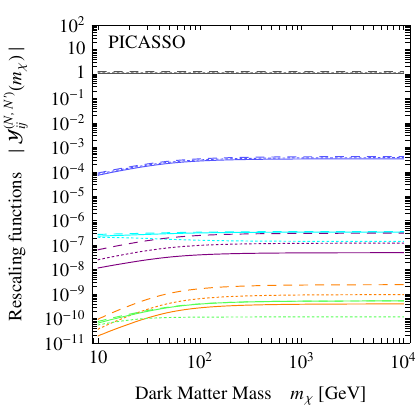} }
\parbox[b]{0.23\linewidth}{
	\raisebox{2.5cm}{\includegraphics[width=0.36\textwidth]{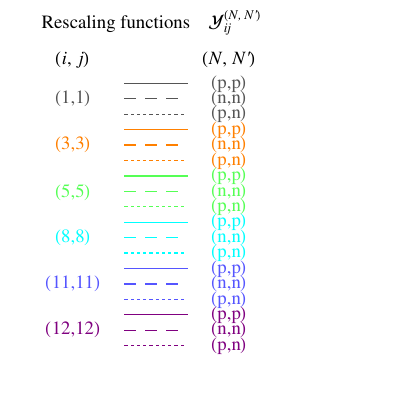}}}
\caption{\em \small \label{fig:Yieqj} 
Absolute value of the {\bfseries rescaling functions $\mathcal{Y}_{i, j}^{(N,N')}(m_\chi)$} for contact interaction and for $i=j$. The ones shown here are those relative to NR operators for which the effective interaction is mostly spin-independent.}
\end{figure}

\begin{figure}[t]
\parbox[b]{0.76\linewidth}{
\includegraphics[width=0.36\textwidth]{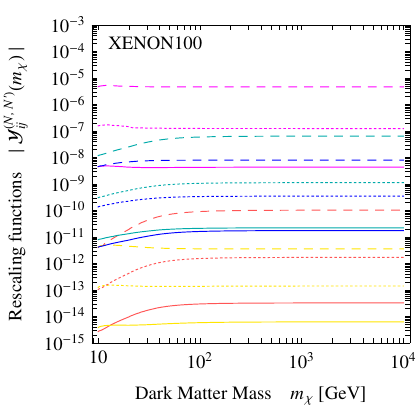} \quad
 	\includegraphics[width=0.36\textwidth]{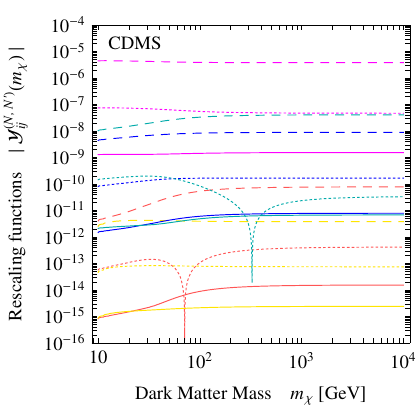}  \\
\includegraphics[width=0.36\textwidth]{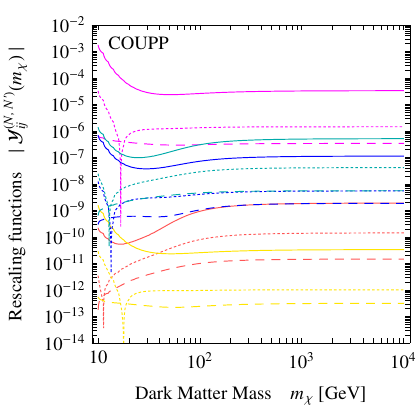} \quad 
 	\includegraphics[width=0.36\textwidth]{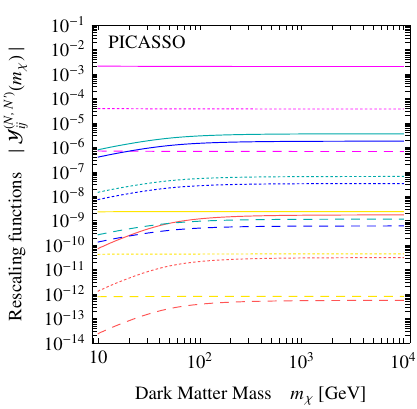} }
\parbox[b]{0.23\linewidth}{
	\raisebox{2.5cm}{\includegraphics[width=0.36\textwidth]{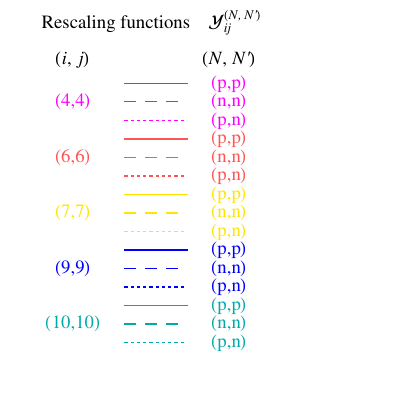}}}
\caption{\em \small \label{fig:Yieqj2} 
Absolute value of the {\bfseries rescaling functions $\mathcal{Y}_{i, j}^{(N,N')}(m_\chi)$} for contact interaction and for $i=j$. The ones shown here are those relative to NR operators for which the effective interaction is mostly spin-dependent.}
\end{figure}

\begin{figure}[t]
\parbox[b]{0.76\linewidth}{
\includegraphics[width=0.36\textwidth]{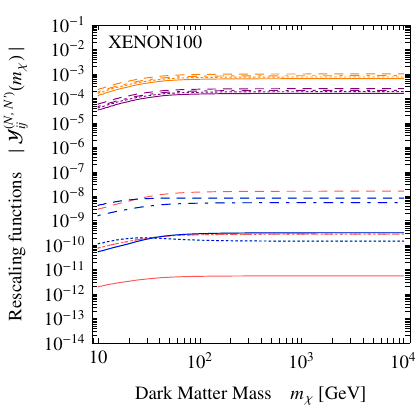} \quad
 	\includegraphics[width=0.36\textwidth]{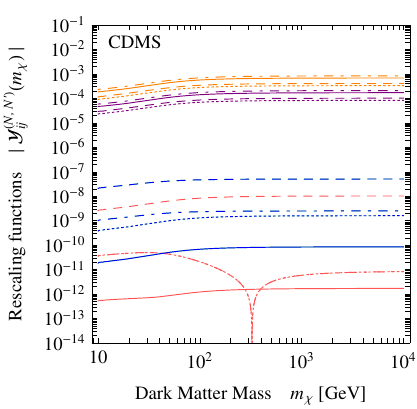}  \\
\includegraphics[width=0.36\textwidth]{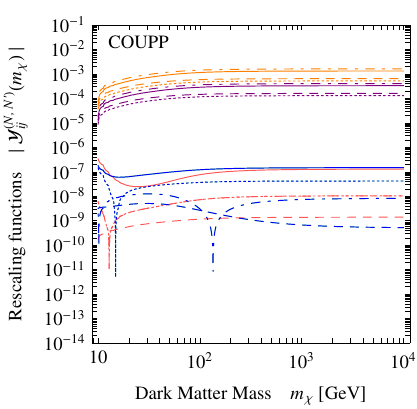} \quad 
 	\includegraphics[width=0.36\textwidth]{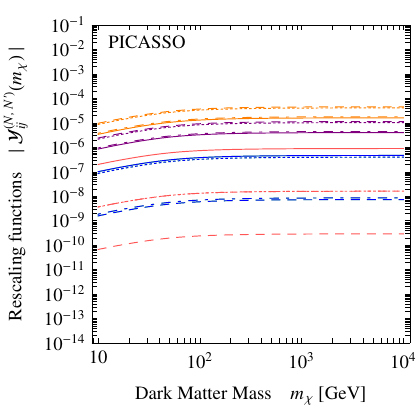} }
\parbox[b]{0.23\linewidth}{
	\raisebox{2.5cm}{\includegraphics[width=0.36\textwidth]{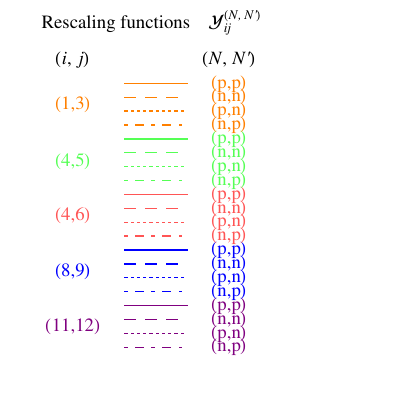}}}
\caption{\em \small \label{fig:Yineqj} 
Absolute value of the {\bfseries rescaling functions $\mathcal{Y}_{i, j}^{(N,N')}(m_\chi)$} for contact interaction, in the case of interference among NR operators with $i\neq j$. Notice that the lines for $(i,j)=(4,5)$ are not visible as they are superimposed with some of the other ones.}
\end{figure}

\begin{figure}[t]
\parbox[b]{0.76\linewidth}{
\includegraphics[width=0.36\textwidth]{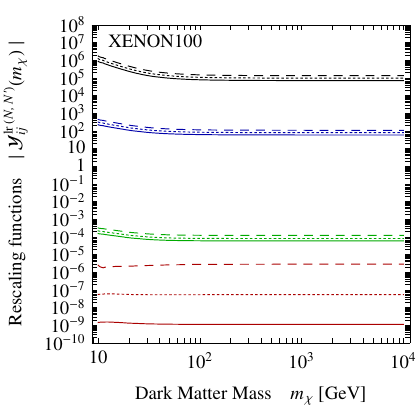} \quad
 	\includegraphics[width=0.36\textwidth]{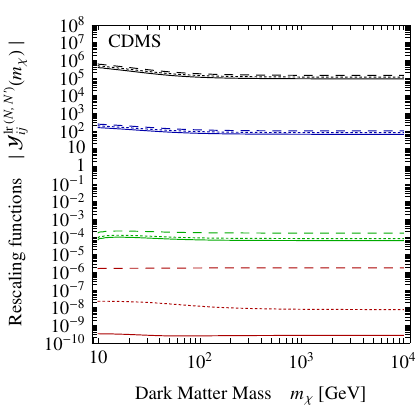}  \\
\includegraphics[width=0.36\textwidth]{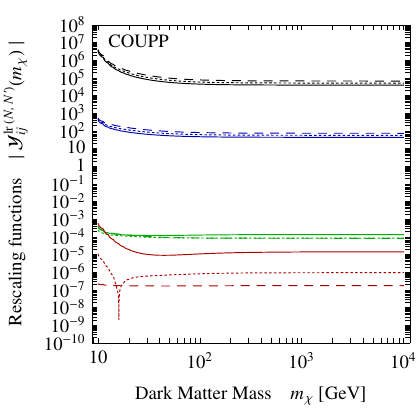} \quad 
 	\includegraphics[width=0.36\textwidth]{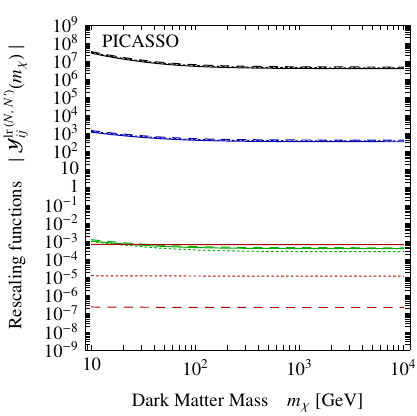} }
\parbox[b]{0.23\linewidth}{
	\raisebox{2.5cm}{\includegraphics[width=0.36\textwidth]{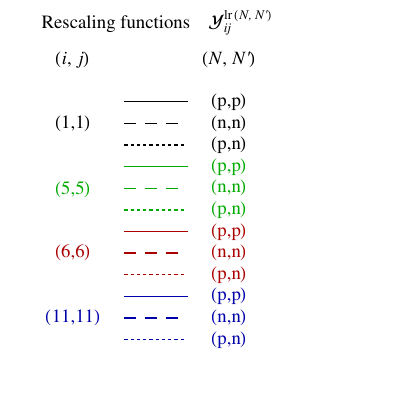}}}
\caption{\em \small \label{fig:YieqjLR} 
Absolute value of the {\bfseries rescaling functions $\mathcal{Y}_{i, j}^{{\rm lr}(N,N')}(m_\chi)$} for long range interaction and for $i=j$.}
\end{figure}

\begin{figure}[t]
\parbox[b]{0.76\linewidth}{
\includegraphics[width=0.36\textwidth]{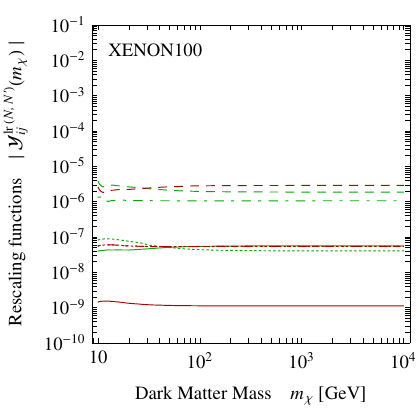} \quad
 	\includegraphics[width=0.36\textwidth]{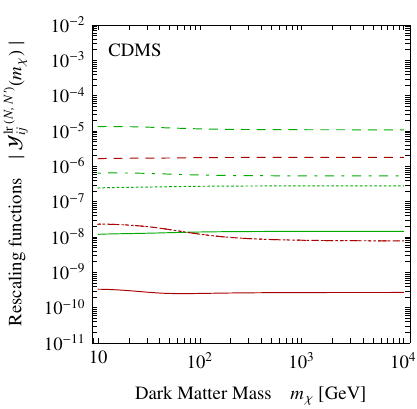}  \\
\includegraphics[width=0.36\textwidth]{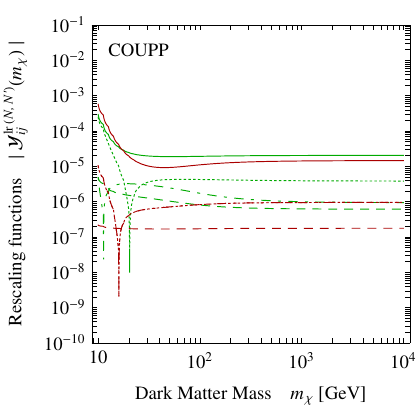} \quad 
 	\includegraphics[width=0.36\textwidth]{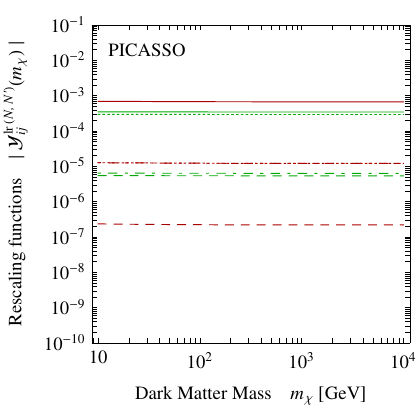} }
\parbox[b]{0.23\linewidth}{
	\raisebox{2.5cm}{\includegraphics[width=0.36\textwidth]{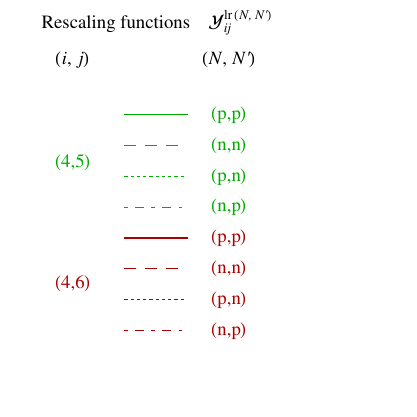}}}
\caption{\em \small \label{fig:YineqjLR} 
Absolute value of the {\bfseries rescaling functions $\mathcal{Y}_{i, j}^{{\rm lr}(N,N')}(m_\chi)$}, in the case of interference among the NR operator $\Op^{\rm NR}_4$ and the NR long range operators $\Op^{\rm lr}_5$ and $\Op^{\rm lr}_6$.}
\end{figure}

\subsection{\XENON}
\label{XENON}
The \XENON\ detector, located in Gran Sasso National Laboratory in Italy, is a two-phase time projection chamber enclosing $62$ kg of target mass. The ratio between the scintillation signal ($S_1$) due to a particle interaction in the liquid xenon, and the subsequent ionization signal ($S_2$), allows for an excellent discrimination of electromagnetic background events. Moreover, the ability to reconstruct the three-dimensional coordinates of each event further enables background reduction by exploiting the self-shielding of liquid xenon in the fiducial volume. Due to their large mass number, Xe nuclei are an excellent target for WIMPs (Weakly Interacting Massive Particles) with spin-independent interactions. However, the unpaired neutron of the $^{129}$Xe and $^{131}$Xe isotopes also makes the experiment sensitive to $n$-WIMP spin-dependent interactions.

In \cite{Aprile:2012nq}, the collaboration reported the results of the last run, a blind analysis with an exposure $w = 34 \times 224.6$ kg$\, \cdot \,$days which yielded no evidence for DM interactions. The two candidate DM events in the pre-defined nuclear recoil energy range of $6.6 - 43.3$ keV$_\text{nr}$ are in fact consistent with the background expectation of $N^{\rm bkg} = 1.0 \pm 0.2$ events.

The experiment detects primary ($S_1$) and secondary ($S_2$) scintillation light, that is converted into photoelectrons (PE) by the photomultiplier tubes (PMT). So what is actually measured is a number of electrons instead of the energy of the event. The steps to convert the theoretical event rate into a quantity that is closer to what the experiment actually measures are illustrated in \cite{Aprile:2011hx}. The event rate in number of photoelectrons $n$ (summing over the Xe isotopes indexed by $T$) is given by
\beq
\frac{\ud R}{\ud n} = \int_0^\infty \ud \ER \, {\rm Poiss}(n | \nu(\ER)) \sum_T \frac{\ud R_T}{\ud \ER} \ ,
\eeq
where $\nu(\ER) = \ER \, \mathcal L_{\rm eff}(\ER) \, L_y \, S_{\rm nr} / S_{\rm ee}$ is the expected number of PE for a given recoil energy and ${\rm Poiss}(n | \nu(\ER))$ describes the  statistical PE distribution. $\mathcal L_{\rm eff}(\ER)$ is the relative scintillation efficiency relating the expected number of $S_1$ photoelectrons to the recoil energy $\ER$, and is provided in figure 1 of \cite{Aprile:2011hi};\footnote{The unknown details of $\mathcal L_{\rm eff}(\ER)$ at low energies, while being crucial for an understanding of the detector performances at low DM masses $m_\chi \sim 10$ GeV, do not affect the \XENON\ bounds on heavier DM \cite{Aprile:2012nq}.}
$L_y = 2.28 \pm 0.04$ PE/keV$_{\rm ee}$ is the light yield and $S_{\rm ee} = 0.58$ and $S_{\rm nr} = 0.95$ are the scintillation quenching factors due to the electric field for electronic and nuclear recoils, respectively. Taking also into account the finite average single-photoelectron resolution $\sigma_{\rm PMT} = 0.5$ PE of the photomultipliers, and the acceptance of the applied cuts $\epsilon(S_1)$ from \cite{Aprile:2012nq}, the resulting $S_1$ spectrum is
\beq
\frac{\ud R}{\ud S_1} = \epsilon(S_1) \sum_{n = 1}^\infty {\rm Gauss}(S_1 | n, \sqrt n \, \sigma_{\rm PMT}) \frac{\ud R}{\ud n} \ .
\eeq
Finally, we compute the total rate as
\begin{multline}
\label{RXENON}
R = \int_{S_1^{\rm min}}^{S_1^{\rm max}} \hspace{-.33cm} \ud S_1 \frac{\ud R}{\ud S_1} =
\int_{S_1^{\rm min}}^{S_1^{\rm max}} \hspace{-.33cm} \ud S_1 \, \epsilon(S_1)
\\
\times \sum_{n = 1}^\infty {\rm Gauss}(S_1 | n, \sqrt n \, \sigma_{\rm PMT}) \int_0^\infty \ud \ER \, {\rm Poiss}(n | \nu(\ER)) \sum_T \frac{\ud R_T}{\ud \ER} \ ,
\end{multline}
where $S_1^{\rm min} = 3$ PE and $S_1^{\rm max} = 30$ PE corresponding to the total energy range $6.6 - 43.3$ keV$_\text{nr}$ used in the analysis.

Substituting the expression for $\ud R_T / \ud \ER$, \Eq{Rate}, within \Eq{RXENON}, and matching with \Eq{Nth} one finally gets the integrated form factors for the \XENON\ experiment:
\begin{multline}
\label{FtildeXe}
\tilde{\mathcal{F}}_{i, j}^{(N,N')}(m_\chi)\rfloor_{\text{\XENON}} = w \sum_T \xi_T
\int_{S_1^{\rm min}}^{S_1^{\rm max}} \hspace{-.33cm} \ud S_1 \, \epsilon(S_1)
\\
\times \sum_{n = 1}^\infty {\rm Gauss}(S_1 | n, \sqrt n \, \sigma_{\rm PMT}) \int_0^\infty \ud \ER \, {\rm Poiss}(n | \nu(\ER)) \, \mathcal{F}_{i, j}^{(N,N')}(\ER, T) \ .
\end{multline}

\subsection{\CDMS}
\label{CDMS}
The Cryogenic Dark Matter Search ({\sc Cdms II}) experiment is located at the Soudan Underground Laboratory in Minnesota. $19$ Germanium and $11$ Silicon detectors measure the energy deposited by incident particles in the form of phonons and ionization. The ratio between these two signals provides event-by-event rejection of electron recoils produced by incident electrons and photons. Due to the reduced ionization collection in the external part of the detectors, electron recoils occurring near the surface (\virg{surface events}) are more likely to be misidentified as nuclear recoils; however, these events can be discriminated and rejected using phonon pulse timing, further lowering the background from electromagnetic events.

In the final data run of the experiment \cite{Ahmed:2009zw}, a blind analysis with an exposure $w = 612$ kg$\,\cdot\,$days has yielded no significant evidence for DM interactions. Only Ge detectors were used;\footnote{Recently the collaboration also presented results based on the Si analysis~\cite{Agnese:2013cvt, Agnese:2013rvf}. We will not consider these here.} due to its large mass number, Ge is well suitable for direct searches of WIMPs with spin-independent interaction. The collaboration observed two candidate events in the signal region of the detector, against an expected background of misidentified surface events of $0.8 \pm 0.1 \text{(stat)} \pm 0.2 \text{(syst)}$; in addition, the expected nuclear recoil background counts $0.04^{+0.04}_{-0.03}$(stat) events from cosmogenic neutrons and $0.03 - 0.06$ events from radiogenic neutrons. We consider therefore a total of $N^{\rm bkg} = 0.9$ background events given by the sum of these independent contributions.

The recoil rate due to DM scatterings in the \CDMS\ detector can be straightforwardly obtained by integrating the theoretical differential rate over the energy window of the detector, taking into account the nuclear recoil efficiency $\epsilon(\ER)$ \cite{Ahmed:2009zw} and summing over germanium's isotopes:
\beq
R = \int_{E^{\rm min}}^{E^{\rm max}} \hspace{-.33cm} \ud \ER \, \epsilon(\ER) \sum_T \frac{\ud R_T}{\ud \ER} \ ,
\eeq
with $E^{\rm min} = 10$ keV$_\text{nr}$ and $E^{\rm max} = 100$ keV$_\text{nr}$. In this case the integrated form factor is given by
\beq
\tilde{\mathcal{F}}_{i, j}^{(N,N')}(m_\chi)\rfloor_{\text{\CDMS}} = w \sum_T \xi_T
\int_{E^{\rm min}}^{E^{\rm max}} \hspace{-.33cm} \ud \ER \, \epsilon(\ER) \, \mathcal{F}_{i, j}^{(N, N')}(\ER, T) \ .
\eeq

\subsection{\COUPP}
\label{COUPP}
The Chicagoland Observatory for Underground Particle Physics (\COUPP) is a 4 kg CF$_3$I bubble chamber operated at SNOLAB in Ontario. Because of its unpaired proton, fluorine provides excellent sensitivity to $p$-WIMP spin-dependent interactions, while iodine enhances the sensitivity to spin-independent interactions due to its large mass number.

Bubble nucleations, triggered by particles entering the liquid in the superheated phase, are recorded both photographically and by pressure and acoustic measurements. With appropriate chamber pressure and temperature, electron recoils due to the abundant gamma-ray and beta-decay backgrounds do not nucleate bubbles; therefore, the main background is constituted by neutrons and alpha decays. Alpha events are discriminated by an acoustic cut, while photographic and pressure measurements are used to identify and reject events occurring outside the fiducial volume as well as events with multiple bubble nucleation.

Even though the experiment is not able to tell the scattering energy on an event-by-event basis, the superheated fluid can be operated at different temperatures, corresponding to different energy thresholds for bubble nucleation. During the last run \cite{Behnke:2012ys}, data were collected in three contiguous data collection periods at temperatures of $39.0^\circ$ C, $36.2^\circ$ C and $33.5^\circ$ C, corresponding to thresholds of $7.8$, $11.0$ and $15.5$ keV$_\text{nr}$, respectively. Unfortunately there is still no precise understanding of how the probability of nucleating a bubble as a function of the energy deposition depends on the threshold. To estimate this uncertainty the collaboration uses two different parametrizations for the nucleation efficiency, fitting the data equally well, thus drawing the uncertainty band visible in figures 6 and 7 of \cite{Behnke:2012ys}.\footnote{Fluorine gives the main contribution to the spin-dependent scattering cross section, while being also important in low energy spin-independent scattering. The large uncertainties on the probability of nucleating a bubble for scattering of a F nucleus in CF$_3$I are therefore responsible for the width of the exclusion bands in figure 6 and (for low DM masses) in figure 7 of \cite{Behnke:2012ys}. At large energies (and therefore for higher DM masses), the bound on the spin-independent interaction is driven instead by iodine, whose behavior is known with better accuracy thus making the band shrink.}

In \cite{Behnke:2012ys}, data obtained for an effective exposure to single recoil events of $437.4$ kg$\,\cdot\,$days (taking into account the $79.1\%$ detection efficiency) were presented. Twenty single nuclear recoil events passing all the analysis cuts were observed over the three energy bins. However, a more accurate analysis of the data led to think that not all these events are genuine single nuclear recoils. Upon implementation of a further time isolation cut, the number of signal events is reduced to thirteen as reported in Table \ref{tabCOUPP}. Due to uncertainties in the neutron background estimation, the collaboration does not attempt any background subtraction and instead treats all the twenty originally observed nuclear recoils as DM events. However, our Test Statistic cannot be applied to the case of zero background, and therefore we adopt the background estimates made by the collaboration, also reported in Table \ref{tabCOUPP}. We enforce the time isolation cut made by the collaboration, for this yields a bound closer to the experimental one.

\begin{table}
\centering
\begin{tabular}{p{2.2cm} | p{3.3cm} | p{2.5cm} | p{2.5cm}}
Nucleation threshold $E_k^\text{thr}$ (keV) & Effective exposure $w_k$ (kg$\,\cdot\,$days) & Observed events $N^\text{obs}_k$ & Expected background $N^\text{bkg}_k$
\\
\hline
$7.8$ & $55.8$ & $2$ & $0.8$
\\
$11.0$ & $70.0$ & $3$ & $0.7$
\\
$15.5$ & $311.7$ & $8$ & $3.0$
\end{tabular}
\caption{\em \small \label{tabCOUPP} {\bfseries Effective exposures, observed counts and expected background} for each data set of the \COUPP\ experiment \cite{Behnke:2012ys}. We only consider the events passing the time isolation cut.}
\end{table}

The rate of DM events in the $k$-$\text{th}$ energy bin, identified by its low energy threshold $E_k^\text{thr}$, is determined by convolving the theoretical differential rate with the target-dependent nucleation efficiency $\mathcal P_T(\ER, E_k^\text{thr})$:
\beq
R_k = \int_{E_k^\text{thr}}^\infty \ud\ER \sum_{T = {\rm C, F, I}} \mathcal P_T(\ER, E_k^\text{thr}) \, \frac{\ud R_T}{\ud\ER} \ .
\eeq
Scattering of iodine in CF$_3$I is known to have a good efficiency, which can be thus taken to be $1$ above threshold, $\mathcal P_\text{I}(\ER, E_k^\text{thr}) = \theta(\ER - E_k^\text{thr})$. For fluorine and carbon the efficiency is uncertain, therefore the collaboration models it as
\beq\label{P-acceptance}
\mathcal P_{\rm C, F}(\ER, E_k^\text{thr}) = 1 - \exp \left[ a_{\rm C, F} \left( 1 - \frac{\ER}{E_k^\text{thr}} \right) \right] \ ,
\eeq
where a fit to the data yields $a_{\rm C, F} = 0.15$. The parameter $a$ determines the rise in the nucleation efficiency with energy (higher $a$ means steeper efficiency). This low value of $a_{\rm F}$ therefore greatly decreases the sensitivity of the detector, especially at low energies. The other parametrization of $\mathcal P(\ER, E_k^\text{thr})$ proposed by the collaboration, consisting of a step function centered at the threshold, yields probably tighter constraints on the DM parameter space. Conservatively, we prefer to use formula \eqref{P-acceptance}, that is also the one used by the \PICASSO\ collaboration (which utilizes fluorine nuclei as well).

The integrated form factors will be then  
\beq\label{FtildeCOUPP}
\tilde{\mathcal{F}}_{i, j}^{(N,N')}(m_\chi)\rfloor_{\text{\COUPP}} = \sum_k w_k \sum_{T = {\rm C, F, I}} \xi_T
\int_{E_k^\text{thr}}^\infty \ud\ER \, \mathcal P_T(\ER, E_k^\text{thr}) \, \mathcal{F}_{i, j}^{(N, N')}(\ER, T) \ .
\eeq
Actually,~\Ref{Fitzpatrick:2012ix} does not provide the form factors for carbon, and therefore we omit its contribution in \Eq{FtildeCOUPP}. We note however that this is an excellent approximation, in that spin-dependent interactions are mainly probed by fluorine, while spin-independent interactions are more sensitive to heavier nuclei such as iodine and fluorine.

\subsection{\PICASSO}
\label{PICASSO}
The \PICASSO\ experiment, located at SNOLAB, searches for Dark Matter using superheated liquid droplets, a variant of the bubble chamber technique. The abundance of fluorine in the target liquid C$_4$F$_{10}$ yields an excellent sensitivity to spin-dependent WIMP interactions with protons, especially at low DM masses due to small fluorine mass and the very low recoil detection threshold.

Similarly to \COUPP, particles which produce low ionization densities such as cosmic muons, gamma rays and beta radiation, become detectable only at high temperatures. They are therefore well separated from strongly ionizing WIMP induced recoils, thus allowing efficient suppression of such backgrounds. Alpha particles feature instead a flat spectrum up to high energy thresholds and constitute therefore the main background.

Of the $32$ detector modules accommodated by \PICASSO, only $10$ were used in the latest data analysis \cite{Archambault:2012pm}. By varying the temperature, the collaboration constructed the spectrum of the particle induced energy depositions exploiting the strict correspondence of detector temperature to threshold energy. After correcting for cut acceptances and dead time, the rates recorded by the modules at each temperature are normalized with respect to the active fluorine mass; as for \COUPP, the contribution of carbon is negligible and we disregard it in the following (the collaborations estimates it as $10\%$ of the total spin-independent cross section \cite{Archambault:2012pm}, while it plays presumably no role in spin-dependent interactions). The counting rates of all detector modules result to be flat in the considered range of temperatures $28^\circ \text{ C } < T < 48^\circ \text{ C}$, corresponding to low threshold energies between $1.7$ and $55$ keV, in contrast with a decreasing expected DM signal. The flat rate can be explained by the presence of alpha emitters in the droplets, whose origin is still uncertain. For this reason, the collaboration does not try to predict the background, rather it fits it to a detector-dependent constant using the data (different modules display different levels of alpha contamination, due to differences in the fabrication processes). 

Since the collaboration does not disclose the data, but only provides averaged quantities, we proceed as follows.
We \virg{virtually} subtract the average background $\overline{R_j} / 8$ from the data points $R_j(T_k)$ of the relative module (indicated by the index $j$) for the eight temperature values $T_k$ probed by the experiment. These rate fluctuations can then be directly compared, for a certain temperature $T_k$, among different modules: averaging over different detectors we then get
\beq\label{PICASSOrate}
\Delta R(T_k) \equiv \frac{\sum_j w_j \left( R_j(T_k) - \overline{R_j} / 8 \right)}{\sum_j w_j} \ ,
\eeq
where $w_j$ is the exposure of the $j^{\rm th}$ module. What one should have used here are actually the exposures relative to the temperature $T_k$, reflecting the time spent by each module at that temperature; since we do not have this information, we assume these exposures to be fixed fractions of the $w_j$'s, independent on the detector.
The $\Delta R(T_k)$ are depicted in figure 5 of \cite{Archambault:2012pm}; the $\overline{R_j}$ and the $w_j$ are given respectively in tables 3 and 1 of \cite{Archambault:2012pm}. All these quantities are reported here in Tab.~\ref{tabPICASSO}.

\begin{table}
\centering
\begin{tabular}{p{2cm} | p{2.5cm} | p{2.5cm}}
Module number $j$ & Average rate $\overline{R_j}$ (cpd/kg) & Exposure $w_j$ (kg$\,\cdot\,$days)
\\
\hline
   71 & $327.6$ & $16.09$ \\
   72 & $134.2$ & $17.69$ \\
 131 & $31.5$ & $10.89$ \\
 134 & $209.6$ & $15.94$ \\
 137 & $69.9$ & $16.33$ \\
 141 & $25.2$ & $13.37$ \\
 144 & $60.8$ & $6.18$ \\
 145 & $31.5$ & $7.83$ \\
 147 & $20.6$ & $6.55$ \\
 148 & $20.0$ & $3.43$ \\
\end{tabular}
\qquad
\raisebox{0.5cm}{
\begin{tabular}{p{2.2cm} | p{3.2cm}}
Threshold energy $E^{\rm thr}_k$ & Rate fluctuations $\Delta R(E^{\rm thr}_k)$
\\
\hline
$1.7$ & $-6.0 \pm 7.1$ \\
$2.9$ & $-0.3 \pm 1.8$ \\
$4.1$ & $1.6 \pm 9.0$ \\
$5.8$ & $-0.2 \pm 9.2$ \\
$6.9$ & $0.05 \pm 1.3$ \\
$16.3$ & $1.4 \pm 1.7$ \\
$38.8$ & $-0.2 \pm 1.7$ \\
$54.8$ & $1.3 \pm 4.7$ \\
\end{tabular}}

\caption{\em \small \label{tabPICASSO} {\bfseries Technical data of the \PICASSO\ analysis}. Left: Exposures and average rates for each of the detectors used in the study. Masses are normalized to the mass of fluorine, \ie kg = kg(F). From tables 1 and 3 of \Ref{Archambault:2012pm}. Right: Low energy thresholds defining the bins used in the analysis and respective rate fluctuations. Adapted from figure 5 of \Ref{Archambault:2012pm}.}
\end{table}

From \Eq{PICASSOrate} one recovers the following formula for the total rate $R_k^{\rm obs}$ measured by the experiment for a given energy threshold $E^{\rm thr}_k$ (corresponding to a temperature $T_k$):
\beq\label{expRatePICASSP}
R_k^{\rm obs} \equiv \frac{1}{w} \sum_j w_j R_j(E^{\rm thr}_k) = \Delta R(E^{\rm thr}_k) + \overline R \ ,
\eeq
where $w \equiv \sum_j w_j = 114.3$ kg$\,\cdot\,$days and $\overline R \equiv \frac{1}{8 w} \sum_j w_j \overline{R_j} = 14.9$ cpd/kg as from Tab.~\ref{tabPICASSO}. The total rates $R_k^{\rm obs}$ can be computed in any given model of DM interactions, adding to the constant background $\overline R$ the DM contribution in the $k$-${\rm th}$ energy bin $R_k^{\rm th}$, and can then be compared with the experimental outcome \Eq{expRatePICASSP} to a chosen confidence level. In analogy with \COUPP,
\beq
R_k^{\rm th} = \int_{E_k^\text{thr}}^\infty \hspace{-0.2cm} \ud\ER \, \mathcal P_{\rm F}(\ER, E_k^\text{thr}) \, \frac{\ud R_{\rm F}}{\ud\ER} \ ,
\eeq
where $\ud R_{\rm F} / \ud\ER$ is the DM scattering rate off fluorine nuclei. $\mathcal P_{\rm F}(\ER, E_k^\text{thr})$ is defined in \Eq{P-acceptance}, whereas the \PICASSO\ collaboration chooses $a_{\rm F} = 5$ while allowing it to vary within $\pm 2.5$ around this central value.

Notice that, since the collaboration does not provide the exposures $w_k$ for the single energy bins, we are not able to determine the total number of events from the rate, which we used in \Eq{Deltachi2Poiss} to explain the statistical procedure. However, the same procedure can be applied to rates rather than number of events. In this regard we can formally define the numbers $N_k^{\rm th}$, $N_k^{\rm bkg}$ and $N_k^{\rm obs}$ as
\begin{align}
N_k^{\rm th} \equiv \tilde{w} R_k^{\rm th} \ ,
&&
N_k^{\rm bkg} \equiv \tilde{w} \overline R \ ,
&&
N_k^{\rm obs} \equiv \tilde{w} R_k^{\rm obs} \ ,
\end{align}
where $\tilde{w}$ is a dummy exposure that can take any value. The integrated form factors will then be
\beq
\tilde{\mathcal{F}}_{i, j}^{(N,N')}(m_\chi)\rfloor_{\text{\PICASSO}} = \tilde{w} \, \xi_{\rm F}
\sum_k
\int_{E_k^\text{thr}}^\infty \ud\ER \, \mathcal P_{\rm F}(\ER, E_k^\text{thr}) \, \mathcal{F}_{i, j}^{(N, N')}(\ER, {\rm F}) \ .
\eeq
The dependence on $\tilde{w}$ will cancel in the functions $\mathcal{Y}_{i, j}^{(N,N')}(m_\chi)$ by their definition in \Eq{Y}.

\section{A dictionary: from quark/gluon-level relativistic effective operators to NR operators}
\label{sec:dictionary}

In this Section we review how to decompose the high-energy effective operators commonly used in model building into the non-relativistic bricks $\Op^\NR_k$ of \Eq{NRoperators}. The starting point are the interaction operators of the DM with quarks and gluons; we review how the step up to the interaction at the nucleon level is performed and then how these are expressed in terms of NR operators, to which the results of Sec.~\ref{sec:formfactors} can be applied straightforwardly.

Incidentally, we will also see explicitly that different high-energy effective operators might have the same non-relativistic form, corresponding thus to the same $\Op^\NR_k$. This can also give rise to interference effects that might significantly lower or enhance the scattering cross section, thus generating phenomenologies that are usually not taken into account by the effective operators analyses when they consider one operator at a time. 

Notice that, while long-distance QCD effects induce energy-dependent corrections to the scattering amplitude, we will only present the matching from quark and gluon level to the nucleon level at lowest order. Next to leading order effects, including two-nucleon interactions \cite{Prezeau:2003sv}, have been studied in the case of scalar interactions in \Ref{Cirigliano:2012pq}, and for spin-dependent (axial-vector) interactions in \Ref{Menendez:2012tm, Klos:2013rwa}.

\subsection{Effective operators for fermion Dark Matter}
\label{EFT4fermionDM}

At dimension six, the effective operators one can construct with a Dirac neutral DM field $\chi$ and quark fields $q$ are
\begin{equation}
\label{DOperators}
\begin{aligned}
\Op^q_1 &= \bar\chi \chi \ \bar q q \ ,
&
\Op^q_2 &= \bar\chi \, i \gamma^5 \chi \ \bar q q \ ,
\\
\Op^q_3 &= \bar\chi \chi \ \bar q \, i \gamma^5 q \ ,
&
\Op^q_4 &= \bar\chi \, i \gamma^5 \chi \ \bar q \, i \gamma^5 q \ ,
\\
\Op^q_5 &= \bar\chi \gamma^\mu \chi \ \bar q \gamma_\mu q \ ,
&
\Op^q_6 &= \bar\chi \gamma^\mu \gamma^5 \chi \ \bar q \gamma_\mu q \ ,
\\
\Op^q_7 &= \bar\chi \gamma^\mu \chi \ \bar q \gamma_\mu \gamma^5 q \ ,
&
\Op^q_8 &= \bar\chi \gamma^\mu \gamma^5 \chi \ \bar q \gamma_\mu \gamma^5 q \ ,
\\
\Op^q_9 &= \bar\chi \, \sigma^{\mu\nu} \chi \ \bar q \, \sigma_{\mu\nu} q \ ,  \qquad \quad
&
\Op^q_{10} &= \bar\chi \, i \, \sigma^{\mu\nu} \gamma^5 \chi \ \bar q \, \sigma_{\mu\nu} q \ ,
\end{aligned}
\end{equation}
where we do not take into account here flavor-violating interactions. Notice that the operators
\beq
\bar\chi \, \sigma^{\mu\nu} \chi \ \bar q \, i \, \sigma_{\mu\nu} \gamma^5 q \ ,
\qquad\qquad
\bar\chi \, i \, \sigma^{\mu\nu} \gamma^5 \chi \ \bar q \, i \, \sigma_{\mu\nu} \gamma^5 q
\eeq
are equal to $\Op^q_{10}$ and $- \Op^q_9$, respectively, by virtue of the identity $i \, \sigma^{\mu\nu} \gamma^5 = - \frac{1}{2} \, \varepsilon^{\mu\nu\rho\tau} \sigma_{\rho\tau}$. For a Majorana DM, only the bilinears $\bar\chi \chi$, $\bar\chi \gamma^5 \chi$ and $\bar\chi \gamma^\mu \gamma^5 \chi$ are non-zero.

Gauge-invariant interaction operators with gluons arise at dimension seven, and are
\begin{equation}
\label{DOperatorsg}
\begin{aligned}
\Op^g_1 &= \frac{\aS}{12 \pi} \ \bar\chi \chi \, G^a_{\mu\nu} G^a_{\mu\nu} \ ,  \qquad \qquad
&
\Op^g_2 &= \frac{\aS}{12 \pi} \ \bar\chi \, i \gamma^5 \chi \, G^a_{\mu\nu} G^a_{\mu\nu} \ , 
\\
\Op^g_3 &= \frac{\aS}{8 \pi} \ \bar\chi \chi \, G^a_{\mu\nu} \tilde{G}^a_{\mu\nu} \ , \rule{0mm}{7mm}
&
\Op^g_4 &= \frac{\aS}{8 \pi} \ \bar\chi \, i \gamma^5 \chi \, G^a_{\mu\nu} \tilde{G}^a_{\mu\nu} \ ,
\end{aligned}
\end{equation}
where $\tilde{G}^a_{\mu\nu} \equiv \varepsilon^{\mu\nu\rho\sigma} G^a_{\rho\sigma}$, and the numerical overall factors have been chosen for later convenience.

The effective Lagrangian at the quark-gluon level is
\beq
\label{Leffqg}
\Lag_\text{eff} =
 \sum_{k = 1}^{10} \sum_q c^q_k \Op^q_k +
 \sum_{k = 1}^{4} c^g_k \Op^g_k \ ,
\eeq
where the $c^q_k$ and $c^g_k$ are real {\it dimensionful} coefficients:\footnote{The $c^q_k$ coefficients are not to be confused with $c^N_k$ coefficients defined below nor with the $\mathfrak{c}_k^N$ coefficients introduced in Eq.~(\ref{Leff}): the first ones are the coefficients in the expansion in terms of quark/gluon level effective operators, the second ones of the expansion in nucleon level operators, the last ones in NR operators.} $c^q_k$ will have dimensions of [mass]$^{-2}$ and $c^g_k$ of [mass]$^{-3}$. As briefly reviewed in Appendix \ref{qg2N}, these operators induce an effective Lagrangian at the nucleon level
\beq
\label{LeffN}
\Lag_\text{eff} = 
 \sum_{k = 1}^{10} \sum_{N = p,n}  c^N_k \Op^N_k  \ ,
\eeq
where the $\Op^N_k$ ($N = p, n$) are
\begin{equation}
\label{DOperators_f}
\begin{aligned}
\Op^N_1 &= \bar\chi \chi \ \bar{N} N \ ,
&
\Op^N_2 &= \bar\chi \, i \gamma^5 \chi \ \bar{N} N \ ,
\\
\Op^N_3 &= \bar\chi \chi \ \bar{N} \, i \gamma^5 N \ ,
&
\Op^N_4 &= \bar\chi \, i \gamma^5 \chi \ \bar{N} \, i \gamma^5 N \ ,
\\
\Op^N_5 &= \bar\chi \gamma^\mu \chi \ \bar{N} \gamma_\mu N \ ,
&
\Op^N_6 &= \bar\chi \gamma^\mu \gamma^5 \chi \ \bar{N} \gamma_\mu N \ ,
\\
\Op^N_7 &= \bar\chi \gamma^\mu \chi \ \bar{N} \gamma_\mu \gamma^5 N \ ,
&
\Op^N_8 &= \bar\chi \gamma^\mu \gamma^5 \chi \ \bar{N} \gamma_\mu \gamma^5 N \ ,
\\
\Op^N_9 &= \bar\chi \, \sigma^{\mu\nu} \chi \ \bar{N} \, \sigma_{\mu\nu} N \ ,
&
\Op^N_{10} &= \bar\chi \, i \, \sigma^{\mu\nu} \gamma^5 \chi \ \bar{N} \, \sigma_{\mu\nu} N \ ,
\end{aligned}
\end{equation}
and we denoted with $N$ the nucleon field. Notice that the gluon operators contribute to the scalar operators $\Op^N_1$, $\Op^N_2$, $\Op^N_3$ and $\Op^N_4$. The couplings are
\begin{subequations}
\label{c^N_k}
\begin{align}
c^N_{1, 2} &= \sum_{q = u, d, s} c^q_{1, 2} \frac{m_N}{m_q} f_{Tq}^{(N)} + \frac{2}{27} f_{TG}^{(N)} \left( \sum_{q = c, b, t} c^q_{1, 2} \frac{m_N}{m_q} - c^g_{1, 2} m_N \right) \ , \label{c^N_k12}
\\
c^N_{3, 4} &= \sum_{q = u, d, s} \frac{m_N}{m_q} \left[ (c^q_{3, 4} - C_{3, 4}) + c^g_{3, 4} \bar{m} \right] \Delta_q^{(N)} \ ,
\\
c^p_{5, 6} &= 2 \, c^u_{5, 6} + c^d_{5, 6} \ ,
\quad
c^n_{5, 6} = c^u_{5, 6} + 2 \, c^d_{5, 6} \ ,
\\
c^N_{7, 8} &= \sum_q c^q_{7, 8} \, \Delta_q^{(N)} \ ,
\\
c^N_{9, 10} &= \sum_q c^q_{9, 10} \, \delta_q^{(N)} \ ,
\end{align}
\end{subequations}
where $C_{3, 4} \equiv \sum_q c^q_{3, 4} \, \bar{m} / m_q$ with $\bar{m} \equiv (1 / m_u + 1 / m_d + 1 / m_s)^{-1}$; the factors $f_{Tq}$, $f_{TG}$, $\Delta_q^{(N)}$ and $\delta_q^{(N)}$ are given in Appendix \ref{qg2N}, to which the reader should refer for the derivation of the equations \eqref{c^N_k} and where reference to the relevant literature is provided. For the quark scalar couplings $c^q_1$ to $c^q_4$ it is usually assumed $c^q_k \propto m_q$, as it would be the case for a DM-quark interaction mediated by a higgs-like particle coupling to the quark masses (we will go back to this point in Sec.~\ref{sec:individualoperators}).

\medskip

To compute the DM-nucleus amplitude, we have now to coherently sum up the interaction amplitude over all nucleons in the nucleus, taking also into account the physics of the bound state. The form factor formalism of \Ref{Fitzpatrick:2012ix} allows to make this for any type of interaction, only we have first to evaluate the matrix elements $_{\rm out}\langle \chi, N | \Op^N_k | \chi, N \rangle_{\rm in}$ and to express them in terms of the non-relativistic operators $\Op^\NR_i$ introduced in \Eq{NRoperators}. To do so, we can expand the solution of the Dirac equation in its non-relativistic limit: in the Weyl or chiral representation for the spinors,
\begin{equation}
\begin{aligned}
u^s (p) =
\begin{pmatrix}
\sqrt{p^\mu \sigma_\mu} \, \xi^s
\\
\sqrt{p^\mu \bar{\sigma}_\mu} \, \xi^s
\end{pmatrix}
&=
\frac{1}{\sqrt{2 (p^0 + m)}}
\begin{pmatrix}
(p^\mu  \sigma_\mu + m) \, \xi^s
\\
(p^\mu \bar{\sigma}_\mu + m) \, \xi^s
\end{pmatrix}
\\
&=
\frac{1}{\sqrt{4 m}}
\begin{pmatrix}
(2 m - \vec{p} \cdot \vec{\sigma}) \, \xi^s
\\
(2 m + \vec{p} \cdot \vec{\sigma}) \, \xi^s
\end{pmatrix}
+ \mathcal{O} (\vec{p}^{\, 2})
\end{aligned}
\end{equation}
where $\sigma^\mu = (\unop, \vec\sigma)$, $\bar\sigma^\mu = (\unop, - \vec\sigma)$ and we approximated $p^\mu = (m, \vec{p}) + \mathcal{O} (\vec{p}^{\, 2})$ in the non-relativistic limit. In this limit we can study the velocity, momentum and spin dependence of the fermion bilinears, when both fermions are on-shell. Up to and including the first order in the three-momenta,
\begin{subequations}
\label{fermionbilinears}
\begin{align}
\bar{u}(p') u(p) &\simeq 2 m \ ,
\\
\bar{u}(p') i \, \gamma^5 u(p) &\simeq 2 i \, \vec{q} \cdot \vec{s} \ ,
\\
\bar{u}(p') \gamma^\mu u(p) &\simeq
\begin{pmatrix}
2m
\\
\vec{P} + 2 i \, \vec{q} \times \vec{s}
\end{pmatrix} \ ,
\\
\bar{u}(p') \gamma^\mu \gamma^5 u(p) &\simeq
\begin{pmatrix}
2 \vec{P} \cdot \vec{s}
\\
4 m \, \vec{s}
\end{pmatrix} \ ,
\\
\bar{u}(p') \sigma^{\mu \nu} u(p) &\simeq
\begin{pmatrix}
0 & i \, \vec{q} - 2 \vec{P} \times \vec{s}
\\
- i \, \vec{q} + 2 \vec{P} \times \vec{s}
&
4 m \, \varepsilon_{i j k} s^k
\end{pmatrix} \ ,
\\
\bar{u}(p') i \, \sigma^{\mu \nu} \gamma^5 u(p) &\simeq
\begin{pmatrix}
0
&
- 4 m \vec{s}
\\
4 m \vec{s}
&
i \, \varepsilon_{i j k} q_k - 2 P_i s^j + 2 P_j s^i
\end{pmatrix} \ ,
\end{align}
\end{subequations}
where $\vec{q} = \vec{p} - \vec{p} \, '$ is the exchanged momentum, and $\vec{P} = \vec{p} + \vec{p} \, '$. The spin operator is defined as $\vec{s} \equiv \xi'^\dagger \frac{\vec{\sigma}}{2} \xi$, where in its absence a $\xi'^\dagger \xi$ is understood.

Finally, when contracting fermionic DM and nucleon bilinears,\footnote{The following expression can be useful in expressing the result as function of only $\vec{q}$ and $\vec{v}^\perp$:
\beq
\frac{\vec P_\chi}{m_\chi} - \frac{\vec P_N}{m_N} = 2 \, \vec{v}^\perp \ ,
\eeq
where $\vec P_\chi$ ($\vec P_N$) is the sum of the initial and final DM (nucleon) momenta.} the following expressions can be derived for the (matrix elements of the) $\Op^N_k$, at leading order in the non-relativistic expansion:
\begin{equation}
\label{fromNtoNR}
\begin{aligned}
\langle \Op^N_1 \rangle = \langle \Op^N_5 \rangle &= 4 m_\chi m_N \Op^\NR_1 \ ,
\\
\langle \Op^N_2 \rangle &= - 4 m_N \Op^\NR_{11} \ ,
\\
\langle \Op^N_3 \rangle &= 4 m_\chi \Op^\NR_{10} \ ,
\\
\langle \Op^N_4 \rangle &= 4 \Op^\NR_6 \ ,
\\
\langle \Op^N_6 \rangle &= 8 m_\chi \left( + m_N \Op^\NR_8 + \Op^\NR_9 \right) \ ,
\\
\langle \Op^N_7 \rangle &= 8 m_N \left( - m_\chi \Op^\NR_7 + \Op^\NR_9 \right) \ ,
\\
\langle \Op^N_8 \rangle = - \frac{1}{2} \langle \Op^N_9 \rangle &= - 16 \, m_\chi m_N \Op^\NR_4 \ ,
\\
\langle \Op^N_{10} \rangle &= 8 \left( m_\chi \Op^\NR_{11} - m_N \Op^\NR_{10} - 4 m_\chi m_N \Op^\NR_{12} \right) \ .
\end{aligned}
\end{equation}
Here we took $\vec{q}$ to be the momentum transferred by the DM to the nucleus. With these substitutions, the effective Lagrangian (\ref{LeffN}) gives origin to the nucleonic matrix element in the form of Eq.~(\ref{Leff}), and the results of Sec.~\ref{sec:formfactors} can be applied straightforwardly.\footnote{Notice that, while in \Ref{Fitzpatrick:2012ix, Fitzpatrick:2012ib} the matrix element is defined as $_{\rm in} \langle \dots \rangle_{\rm out}$, we use $_{\rm out} \langle \dots \rangle_{\rm in}$ instead. For this reason we expect to get a minus sign for each power of $q$ occurring in the NR operators. However, we use an opposite convention for $\vec{q}$ so that the final result will be the same.}

\medskip

From \Eq{fromNtoNR} one can now see clearly that $\Op^N_1$ and $\Op^N_5$ correspond to the same non-relativistic operator, and so do $\Op^N_8$ and $\Op^N_9$. They are therefore indistinguishable by direct detection experiments alone. This also means that a bound computed on one of these operators is identical to the bound computed on the other, a difference arising only if different coefficients are chosen for the two in the Lagrangian. Furthermore, if a model features both operators, strong cancellations or enhancements of the scattering cross section might arise due to their quantum interference. The simplest case in which this could happen is a theory where the DM interacts with quarks via a scalar {\em and} a vector exchange, thus producing both $\Op^N_1$ and $\Op^N_5$.

\subsection{Effective operators for scalar Dark Matter}
\label{EFT4scalarDM}

For a scalar DM, the following contact operators for interaction with quarks are possible:
\begin{equation}
\label{DOperatorsS}
\begin{aligned}
\Op^q_1 &= \phi^* \phi \ \bar q q \ , 
&
\Op^q_2 &= \phi^* \phi \  \bar q \, i \gamma^5 q \ ,
\\
\Op^q_3 &= i \, (\phi^* \overleftrightarrow{\partial_\mu} \phi) \ \bar q \gamma^\mu q \ , \qquad
&
\Op^q_4 &= i \, (\phi^* \overleftrightarrow{\partial_\mu} \phi) \ \bar q \gamma^\mu \gamma^5 q \ .
\end{aligned}
\end{equation}
The operators in the first line are dimension five, while the others have dimension six and vanish identically for a real field $\phi$. The operator $\partial_\mu (\phi^* \phi) \ \bar q \gamma^\mu q$ does not contribute to processes where the quarks are external because it is proportional (after integration by parts) to the divergence of the conserved current $\bar q \gamma^\mu q$, that vanishes by virtue of the equations of motion. The operator $\partial_\mu (\phi^* \phi) \ \bar q \gamma^\mu \gamma^5 q$, instead, reduces to $2 m_q \, \phi^* \phi \ \bar q \, i \gamma^5 q$ upon integration by parts and application of the equations of motion; it is therefore equivalent to $\Op^q_2$, the only difference being the quark mass factor that can be however absorbed in the arbitrary Lagrangian coefficient.
\\
At dimension six also the terms
\begin{align}
\phi^* \phi \ \bar q \, i \overleftrightarrow{\slashed{D}} q \ ,
& & &
\phi^* \phi \ \bar q \, i \overleftrightarrow{\slashed{D}} \gamma^5 q
\end{align}
exist, where $\overleftrightarrow{D_\mu} = \frac{1}{2} \overleftrightarrow{\partial_\mu} - i e Q_q A_\mu - i g_\text{s} T^a_q g^a_\mu$ is the (\virg{hermitianized}) quark covariant derivative. We do not take into account these operators here.

The most relevant gauge-invariant interaction operators with gluons have dimension six and are
\begin{align}
\label{DOperatorsgS}
\Op^g_1 &= \frac{\aS}{12 \pi} \ \phi^* \phi \ G^a_{\mu\nu} G^a_{\mu\nu} \ ,
&
\Op^g_2 &= \frac{\aS}{8 \pi} \ \phi^* \phi \ G^a_{\mu\nu} \tilde{G}^a_{\mu\nu} \ ,
\end{align}
where the numerical factors have been chosen for later convenience.

The effective Lagrangian at the quark-gluon level is therefore
\beq
\label{LeffqgS}
\Lag_\text{eff} =
 \sum_{k = 1}^4 \sum_q c^q_k \Op^q_k +
 \sum_{k = 1,2} c^g_k \Op^g_k \ ,
\eeq
where the $c^q_n$ and $c^g_n$ are real {\it dimensionful} coefficients: $c^q_k$ for $k=1,2$ will have dimension of [mass]$^{-1}$, $c^q_k$ for $k=3,4$ dimension of [mass]$^{-2}$ and $c^g_k$ of [mass]$^{-3}$. 
This induces the effective Lagrangian at the nucleon level
\beq
\label{LeffNS}
\Lag_\text{eff} = 
 \sum_{k = 1}^{4} \sum_{N = p,n}  c^N_k \Op^N_k  \ ,
\eeq
where the $\Op^N_k$ ($N = p, n$) are
\begin{equation}
\label{DOperators_s}
\begin{aligned}
\Op^N_1 &= \phi^* \phi \ \bar{N} N \ ,
&
\Op^N_2 &= \phi^* \phi \  \bar{N} \, i \gamma^5 N \ ,
\\
\Op^N_3 &= i \, (\phi^* \overleftrightarrow{\partial_\mu} \phi) \ \bar{N} \gamma^\mu N \ , \qquad
&
\Op^N_4 &= i \, (\phi^* \overleftrightarrow{\partial_\mu} \phi) \ \bar{N} \gamma^\mu \gamma^5 N \ ,
\end{aligned}
\end{equation}
and the couplings are, in analogy with the fermion DM case,
\begin{subequations}
\begin{align}
c^N_1 &= \sum_{q = u, d, s} c^q_1 \frac{m_N}{m_q} f_{Tq}^{(N)} + \frac{2}{27} f_{TG}^{(N)} \left( \sum_{q = c, b, t} c^q_1 \frac{m_N}{m_q} - c^g_1 \, m_N \right) \ ,
\\
c^N_2 &= \sum_{q = u, d, s} \frac{m_N}{m_q} \left[ (c^q_2 - C_2) + c^g_2 \, \bar{m} \right] \Delta_q^{(N)} \ ,
\\
c^p_3 &= 2 \, c^u_3 + c^d_3 \ ,
\quad
c^n_3 = c^u_3 + 2 \, c^d_3 \ ,
\\
c^N_4 &= \sum_q c^q_4 \, \Delta_q^{(N)} \ .
\end{align}
\end{subequations}
We again point to \App{qg2N} for the derivation of these formulas and for reference to the relevant literature.
\\
As we did in \Sec{EFT4fermionDM} we can now match the (matrix element of the) relativistic operators \eqref{DOperators_s} to the non-relativistic ones \eqref{NRoperators}, at the leading order in the non-relativistic expansion, obtaining
\begin{equation}
\begin{aligned}
\label{fromNtoNRS}
\langle \Op^N_1 \rangle &= 2 m_N \Op^\text{NR}_1 \ ,
\\
\langle \Op^N_2 \rangle &= 2 \Op^\text{NR}_{10} \ ,
\\
\langle \Op^N_3 \rangle &= 4  m_\chi m_N \Op^\text{NR}_1 \ ,
\\
\langle \Op^N_4 \rangle &= - 8 m_\chi m_N \Op^\text{NR}_7 \ .
\end{aligned}
\end{equation}

\subsection{Long range interactions}

While contact interactions as the ones described above are either independent on the exchanged momentum or momentum-suppressed, long range interactions are enhanced at small momentum transfer. These arise from the exchange of a massless mediator, whose propagator is responsible for the enhancement. Since these interactions feature quite different spectra with respect to the ones seen so far, it is worth discussing also this case. For definiteness we take into account here interactions of the DM with the Standard Model photon, that have already been studied in the context of DM direct searches (see \eg \cite{Pospelov:2000bq, Sigurdson:2004zp, Foot:2008nw, Barger:2010gv, Chang:2010en, Ho:2012bg, Fornengo:2011sz, DelNobile:2012tx} and references therein). Notice that interactions with gluons are not of long range type, as at the low energies relevant for direct detection the interaction is effectively with the entire nucleon (see above).

The most relevant gauge-invariant interactions a fermionic Dirac DM $\chi$ can have with photons are
\begin{align}
\label{LRopsferm}
\Op_\text{C} &= Q_\chi e \ \bar\chi \gamma^\mu \chi \, A_\mu \ ,
& \text{(millicharged DM)}
\\
\Op_\text{M} &= \frac{\mu_\chi}{2} \, \bar\chi \, \sigma^{\mu\nu} \chi \, F_{\mu\nu} \ ,
& \text{(anomalous DM magnetic moment)}
\\
\label{LRopsferm3} \Op_\text{E} &= \frac{d_\chi}{2} \, i \, \bar\chi \, \sigma^{\mu\nu} \gamma^5 \chi \, F_{\mu\nu} \ .
& \text{(DM electric dipole moment)}
\end{align}
The dimensionless constant $Q_\chi$ is the DM electric charge in units of $e$, whereas the parameters $\mu_\chi$ and $d_\chi$, with dimension [mass]$^{-1}$, are the magnetic and electric dipole moments (in units of $e \cdot \, \text{cm}$) of $\chi$, respectively. All these interactions vanish identically for Majorana DM particles; the lowest electromagnetic moment allowed for a Majorana particle is the anapole moment, $\bar \chi \, \gamma^\mu \gamma^5 \chi \, \partial^\nu F_{\mu\nu}$, but this does not lead to a long range interaction and therefore we will not consider it here \cite{Ho:2012bg}. These interactions generate the following non-relativistic nucleonic matrix element \eqref{Leff}, when the photon is to mediate the interaction with nucleons $N = p, n$ (see also \cite{Fitzpatrick:2012ib}):
\begin{align}
\Mel^N_\text{C} &= 4 \, e^2 Q_\chi Q_N \, m_\chi m_N \, \Op^\text{lr}_1 \ , \label{LagC}
\\
\Mel^N_\text{M} &= 2 \, e \mu_\chi \left[ Q_N \, m_N \, \Op^\NR_1 + 4 \, Q_N \, m_\chi m_N \, \Op^\text{lr}_5 + 2 \, g_N \, m_\chi \left( \Op^\NR_4 - \Op^\text{lr}_6 \right) \right] \ , \rule{0mm}{6mm}
\\
\Mel^N_\text{E} &= 8 \, e d_\chi Q_N \, m_\chi m_N \, \Op^\text{lr}_{11} \ , \label{LagC3} \rule{0mm}{6mm}
\end{align}
where $Q_p = 1$, $Q_n = 0$ are the nucleons electric charge while $g_p = 5.59$, $g_n = -3.83$ are their $g$-factors. The operators $\Op^\text{lr}_k \equiv \Op^\NR_k / q^2$ have been defined in Sec.~\ref{sec:formfactors}. Consequently, we define appropriate form factors for the long range case, that take into account this dependence on $q$; from these, the relative rescaling functions $\mathcal{Y}_{i, j}^{\text{lr}(N,N')}(m_\chi)$ are then derived. In this way one is not concerned with any momentum dependence of the coefficients in \eqref{Leff}; for example, for millicharged DM \eqref{LagC} we have $\mathfrak{c}^p_1 = 4 \, Q_\chi \, m_\chi m_N$ and $0$ for all the other coefficients.

\spazio

For a scalar DM $\phi$, the only relevant interaction operator with the photon displaying long range behavior is the one coming from the kinetic term, in case the DM has a small electric charge:
\beq\label{phiEM}
\Mel^N_\text{C} = Q_\phi e \, i \, (\phi^* \overleftrightarrow{\partial_\mu} \phi) A^\mu \ .
\eeq
For this operator, the effective interaction Lagrangian with nucleons is the same as \Eq{LagC}, with the obvious substitution $Q_\chi \to Q_\phi$.

Other possible interaction operators of $\phi$ with photons, arising at dimension six, are
\beq\label{phiEMeff}
i \, (\phi^* \overleftrightarrow{\partial_\mu} \phi) \, \partial_\nu F^{\mu\nu} \ ,
\qquad\qquad
\phi^* \phi \, F^{\mu\nu} F_{\mu\nu} \ ,
\qquad\qquad
\phi^* \phi \, F^{\mu\nu} \tilde{F}_{\mu\nu} \ ,
\eeq
with $\tilde{F}^{\mu\nu} \equiv \varepsilon^{\mu\nu\rho\sigma} F_{\rho\sigma}$ (see \eg \Ref{DelNobile:2011uf}). The terms $\partial^\mu \partial^\nu (\phi^* \phi) \overset{\mbox{\tiny$(\sim)$}}{F}_{\!\! \mu\nu}$ and $i \, (\phi^* \overleftrightarrow{\partial_\mu} \phi) \, \partial_\nu \tilde{F}^{\mu\nu}$, vanish identically because of the antisymmetricity of $F_{\mu\nu}$ and $\varepsilon_{\mu\nu\rho\sigma}$. However, these interactions are not of long range type (see \eg \cite{DelNobile:2011je, Frandsen:2012db}) and therefore we don't include them here.

\section{Summary and explicit examples}
\label{sec:examples}

In this last Section we summarize  the steps one can take without even reading the rest of the paper to be able to use our results and start setting bounds on his/her favorite DM candidate.

\begin{enumerate}
\item[1a.] \label{step1} Compute the amplitude for the relevant scattering process in your favorite DM model. At the end of the day it should consist of a linear combination of quark- and gluon-level high-energy operators. The combination will  include, but will not generally be limited to, the operators of Eq.~(\ref{DOperators}) and (\ref{DOperatorsg}), for fermionic DM, or Eq.~(\ref{DOperatorsS}) and (\ref{DOperatorsgS}) for scalar DM (\ie the amplitude should be in the form of Eq.~(\ref{Leffqg}) or (\ref{LeffqgS})). The amplitude can also contain the long range operators of Eq.~(\ref{LRopsferm}) to (\ref{LRopsferm3}). \\
The coefficients of the operators (the $c_k^{q,g}$ of the contact effective operators or $Q_\chi$, $\mu_\chi$, $d_\chi$) encapsulate the particle physics details of the model, including the unknown parameter(s) $\lambda$ on which you wish to derive a bound. 

\item[1b.] Dress up the quark- and gluon-level operators to the nucleon level in the standard way. For the effective operators, we review this process in Sec.~\ref{sec:dictionary} and Appendix \ref{qg2N}. One obtains an effective Lagrangian in the form of Eq.~(\ref{LeffN}) or (\ref{LeffNS}). The coefficients $c_k^N$ now encapsulate the particle physics (including $\lambda$) and the internal nucleon dynamics of quarks and gluons.

\item[1c.] Reduce to the NR limit as illustrated in Sec.~\ref{sec:dictionary}, in particular using the dictionary provided by Eq.~(\ref{fromNtoNR}), (\ref{fromNtoNRS}) or (\ref{LagC})-(\ref{LagC3}) and (\ref{phiEM}). Matching the terms obtained in this expansion with the NR operators in \Eq{NRoperators} will allow to identify the $\mathfrak{c}^N_i(\lambda, m_\chi)$ as the respective coefficients, as in \Eq{Leff}.

\item[2a.] For each experiment, plug in the $\mathfrak{c}^N_i(\lambda, m_\chi)$ in Eq.~(\ref{sol1}), using the interpolated functions $\mathcal{Y}_{i, j}^{(N,N')}(m_\chi)$ that we provide on \href{http://www.marcocirelli.net/NRopsDD.html}{the website}. This provides you with an expression for $\lambda_{\rm B}$, implicit in the unknown parameter $\lambda$. 

\item[2b.] Plug the above expression for $\lambda_{\rm B}$ in the Mathematica interpolated function ${\rm TS}(\lambda_\text{B}, m_\chi)$ of the benchmark model, that we provide on \href{http://www.marcocirelli.net/NRopsDD.html}{the website}. At this point you possess a ${\rm TS}(\lambda, m_\chi)$ function and you can derive a bound on $\lambda$ at the desired confidence level, \eg draw a contour plot of ${\rm TS} = 2.71$ for a 90\% CL.
\end{enumerate}

\noindent We also present a few explicit examples. The first one is the simplest possible case one can imagine: considering one effective high-energy operator at a time. The second example aims to make contact between our formalism and the usual bounds in terms of spin-independent (SI) and spin-dependent (SD) cross sections. These, together with bounds on millicharged DM and DM with magnetic moment interaction, are also illustrated in the sample Mathematica notebook provided on \href{http://www.marcocirelli.net/NRopsDD.html}{the website}. The third example is a more involved, but well defined, framework inspired by Minimal Dark Matter, in which the DM-ordinary matter scattering involves a combination of several (high-energy) operators with non trivial coefficients.

\subsection{Bounds on individual relativistic effective operators}
\label{sec:individualoperators}

\begin{figure}[p]
\begin{centering}
\includegraphics[width=.27\textwidth]{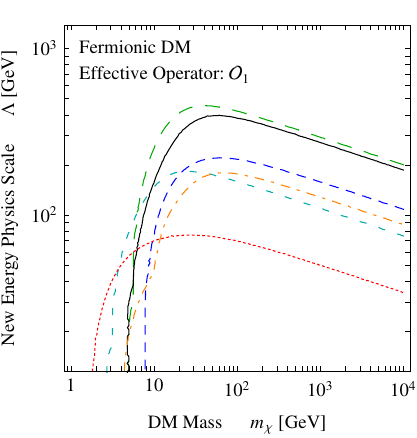}\quad
\includegraphics[width=.27\textwidth]{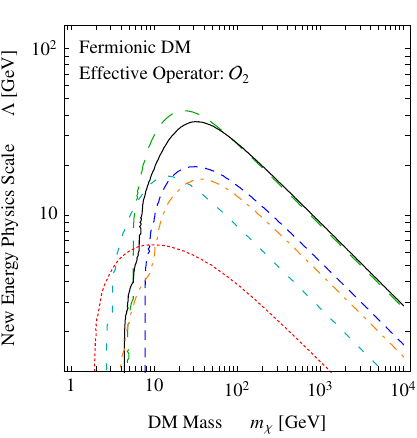}\quad
\includegraphics[width=.27\textwidth]{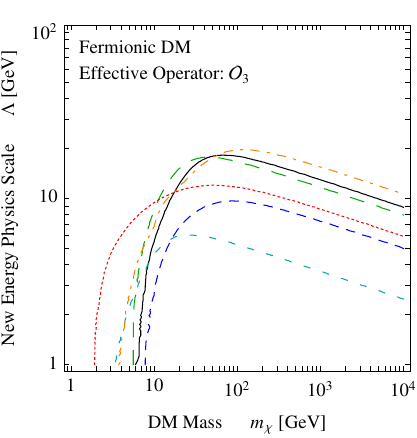} \\[0.1cm]
\includegraphics[width=.27\textwidth]{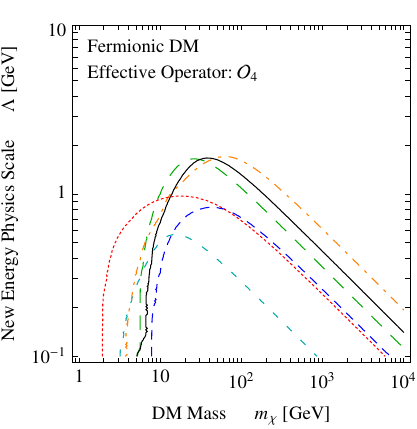}\quad
\includegraphics[width=.27\textwidth]{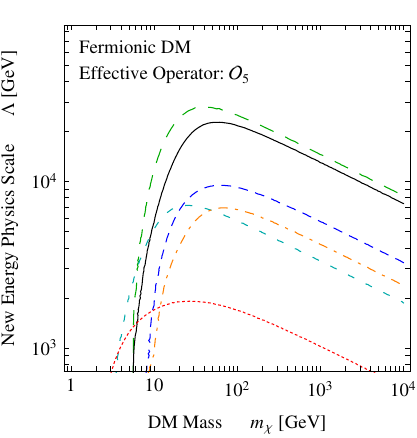}\quad
\includegraphics[width=.27\textwidth]{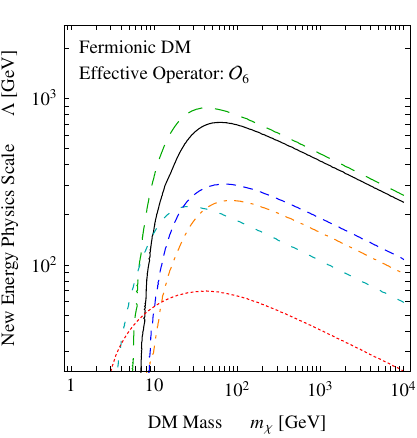} \\[0.1cm]
\includegraphics[width=.27\textwidth]{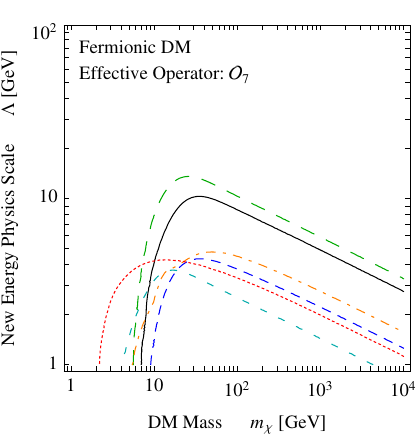}\quad
\includegraphics[width=.27\textwidth]{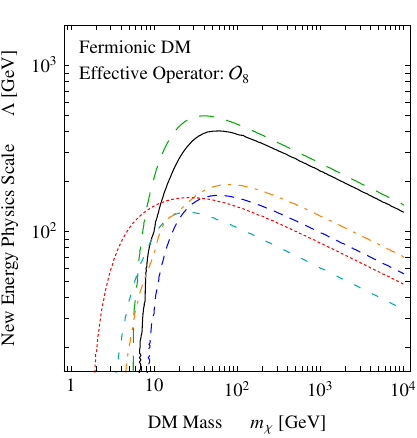}\quad
\includegraphics[width=.27\textwidth]{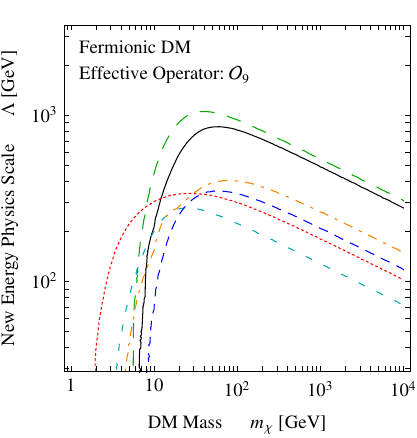} \\[0.1cm]
\includegraphics[width=.27\textwidth]{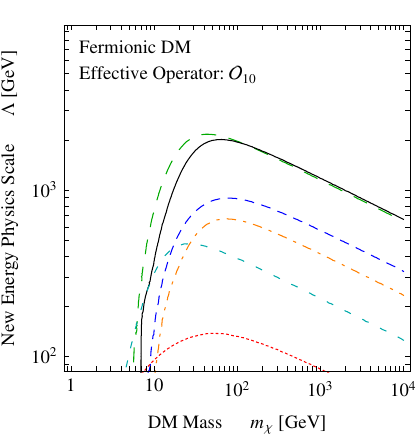}\quad
\includegraphics[width=.27\textwidth]{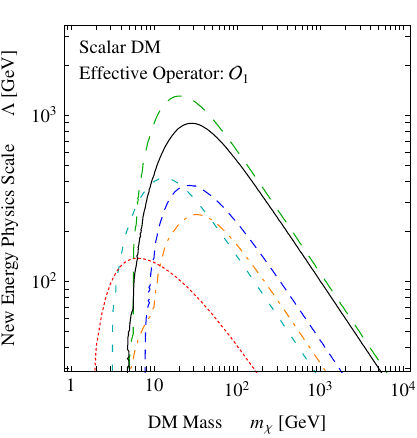}\quad
\includegraphics[width=.27\textwidth]{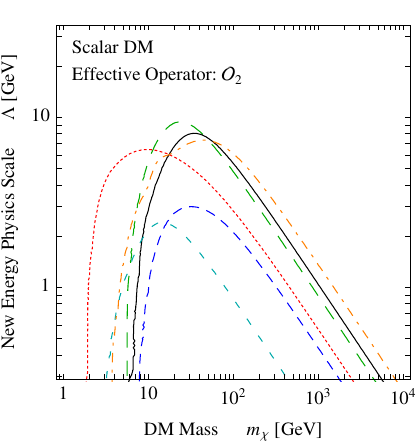} \\[0.1cm]
\includegraphics[width=.27\textwidth]{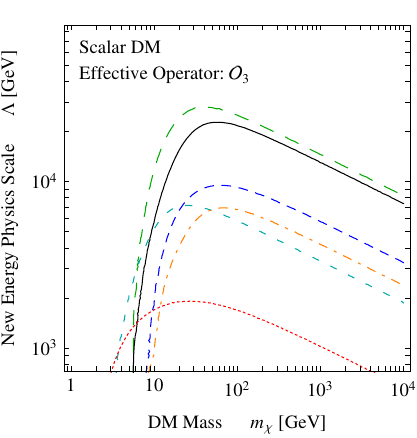}\quad
\includegraphics[width=.27\textwidth]{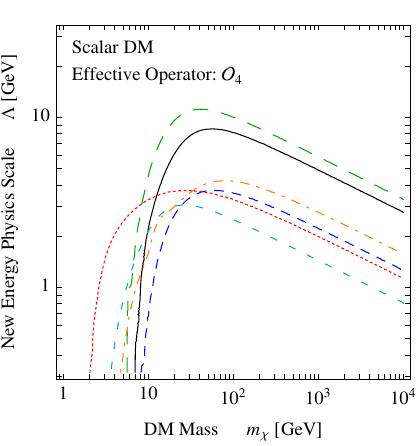}\quad 
\includegraphics[width=.27\textwidth]{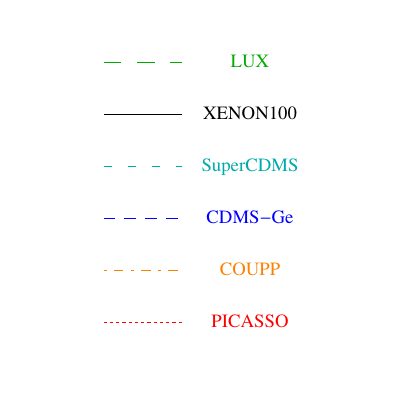} 
\vspace{-0.3cm} \caption{\em \small \label{fig:boundsindividual} 
Bounds on {\bfseries individual} contact {\bfseries effective operators}, all at 90\% CL.}
\end{centering}
\end{figure}

The simplest application of our formalism is the idealized case in which one considers one high-energy contact operator at a time. This is what actually is routinely done in papers employing the effective operator formalism for DM searches (see \eg \Ref{Goodman:2010ku, Fox:2011fx, Cheung:2012gi, Zheng:2010js, Yu:2011by}). We use this simple case to explicitate the steps outlined above in a concrete example. 
For instance, if we study the case of the $k=1$ quark operator for fermionic DM
\begin{equation}
\label{Lagindividual}
\Lag_{\rm eff} \equiv \frac{m_q}{\Lambda^3} \Op_1^q \qquad \Longrightarrow \quad c_1^q = \frac{m_q}{\Lambda^3}, \quad c_{k\neq 1}^q = c_k^g \equiv 0 \ ,
\end{equation}
where $\Lambda$ plays the role of the sole unknown parameter and, as usual, it represents the scale of the ultraviolet physics which has been integrated out. A possible overall numerical factor is conventionally assumed to be 1 or equivalently to be absorbed in the unknown scale $\Lambda$. Next, as dictated by Eq.~(\ref{c^N_k12})
\begin{equation}
c_1^N = \left( \sum_{u,d,s} f^{(N)}_{Tq} + \frac{2}{9}f_{TG}^{(N)} \right) \frac{m_N}{\Lambda^3} \ .
\end{equation}
Thus, following Eq.~(\ref{fromNtoNR}) 
\begin{equation}
\mathfrak{c}_1^N = 4 m_\chi \left( \sum_{u,d,s} f^{(N)}_{Tq} + \frac{2}{9}f_{TG}^{(N)} \right) \frac{m_N^2}{\Lambda^3} \ .
\end{equation}
It is now straightforward to apply Eq.~(\ref{sol1}) and obtain
\begin{equation}
\lambda_{\rm B}^2 =\frac{16 \, m_\chi^2}{\Lambda^6} \sum_{N,N'=p,n} m^2_N \left( \sum_{u,d,s} f^{(N)}_{Tq} + \frac{2}{9}f_{TG}^{(N)} \right) m^2_{N'} \left( \sum_{u,d,s} f^{(N')}_{Tq} + \frac{2}{9}f_{TG}^{(N')} \right) \mathcal{Y}_{1,1}^{(N,N')}(m_\chi) \ .
\end{equation}
Now by explicitly plugging this expression in the ${\rm TS}(\lambda_{\rm B}, m_\chi)$ function for each experiment, one has the ${\rm TS}(\Lambda, m_\chi)$ function which is needed to impose a bound on $\Lambda$. In Fig.~\ref{fig:boundsindividual}, top left panel, we show the result using the values of the $f^{(N)}_{Tq}$ quoted in \Ref{Gondolo:2004sc} (see Table \ref{nuclearities} in Appendix \ref{qg2N}).

We repeat the exercise for all the high-energy contact operators in Eq.~(\ref{DOperators}) and Eq.~(\ref{DOperatorsS}), showing the results in Fig.~\ref{fig:boundsindividual}. For $\Op_{1,2,3,4}^q$ for fermionic DM we assume coefficients proportional to $m_q/\Lambda^3$, like in (\ref{Lagindividual}) as it is customary (since this applies in the case in which the mediator particle couples to quarks like the higgs boson does). Analogously, we assume coefficients proportional to $m_q/\Lambda^2$ for scalar DM operators $\Op_{1,2}^q$.

\subsection{`Usual' SI and SD interactions}
As another example, we discuss here how to recover, within our formalism, the usual spin-independent and spin-dependent bounds presented by the experiments.

As we saw in \Sec{sec:formfactors}, there are different non-relativistic interactions that lead to a detector response which is independent  on the nucleus spin or dependent on the nucleus spin. When experiments present results in terms of `spin-independent' or `spin-dependent' scattering cross section, however, they implicitly assume the lowest order interactions in the non-relativistic expansion, namely $\Op_1^\NR$ for the SI case and $\Op_4^\NR$ for the SD. As specified in Eq.~(\ref{fromNtoNR}), these interactions correspond to the relativistic operators $\Op^N_1 = \bar\chi \chi \, \bar{N} N$ and $\Op^N_8 = \bar\chi \gamma^\mu \gamma^5 \chi \, \bar{N} \gamma_\mu \gamma^5 N$ for fermion DM in Eq.~(\ref{DOperators_f}), respectively. While also $\Op^N_5$ and $\Op^N_9$ lead to $\Op_1^\NR$ and $\Op_4^\NR$, respectively, and could be used for the SI and SD analyses in place of $\Op^N_1$ and $\Op^N_8$, the former ones have been historically considered first (probably as they are the operators relevant in describing elastic scattering of the supersymmetric neutralino).
The effective DM-nucleon Lagrangians are
\begin{align}
\label{LagSI}
\Lag_{\rm SI}^N &= \lambda_{\rm SI} \, \Op^N_1 \ ,
\\
\label{LagSD}
\Lag_{\rm SD}^N &= \lambda_{\rm SD} \, \Op^N_8 \ .
\end{align}
From this, using \Eq{fromNtoNR} to convert the operators to their non-relativistic limit expression, one immediately obtains 
\begin{align}
\label{cfrakSI}
\mathfrak{c}^N_1(\lambda_{\rm SI}, m_\chi) &= 4\, \lambda_{\rm SI} \, m_\chi m_N \ ,
\\
\label{cfrakSD}
\mathfrak{c}^N_4(\lambda_{\rm SD}, m_\chi) &= - 16\, \lambda_{\rm SD}  \, m_\chi m_N \ .
\end{align}
At this point a bound can be set on $\lambda_{\rm SI}$ and $\lambda_{\rm SD}$ using Eq.~(\ref{sol1}). Rather than presenting this kind of bounds, let us make contact with the usual physical cross sections. 

\medskip

For SI scattering, DM-proton and DM-neutron couplings are customarily assumed to be equal, and so are the two cross sections (neglecting the proton-neutron mass difference). These read $\sigma_p^{\rm SI} \equiv \sigma_n^{\rm SI} = \lambda_{\rm SI}^2 \, \mu_N^2 / \pi$, where $\mu_N$ is the DM-nucleon reduced mass. Therefore, to determine a bound on $\sigma_p$ from the usual SI interaction within the framework proposed in this work, one just needs to use $\lambda_{\rm SI}^2 = \pi \sigma_p^{\rm SI} / \mu_N^2$ in Eq.~(\ref{sol1}), together with the value of $\mathfrak{c}^N_1(m_\chi)$ given in \Eq{cfrakSI} and the $\mathcal{Y}_{1,1}^{(N,N')}(m_\chi)$ functions provided in \href{http://www.marcocirelli.net/NRopsDD.html}{the website}. That is, use
\beq
\lambda_\text{B}(m_\chi)^2 = \sigma_p^{\rm SI} \cdot 16 \pi m_\chi^2 \frac{m_N^2}{\mu_N^2} \left( \mathcal{Y}_{1,1}^{(p,p)}(m_\chi) + 2 \, \mathcal{Y}_{1,1}^{(p,n)}(m_\chi) + \mathcal{Y}_{1,1}^{(n,n)}(m_\chi) \right)
\eeq
in the Mathematica interpolated function ${\rm TS}(\lambda_{\rm B}, m_\chi)$ provided on \href{http://www.marcocirelli.net/NRopsDD.html}{the website}. Notice that, exactly as for the original form factors of \Ref{Fitzpatrick:2012ix}, one has $\mathcal{Y}_{i, j}^{(N,N')} = \mathcal{Y}_{j, i}^{(N',N)}$.

\medskip

In the SD case, contrarily to what happens for the SI interaction, the DM-$p$ and DM-$n$ cross sections are usually considered separately, in the assumption that only protons or neutrons contribute, respectively. The cross section on free nucleons is given by $\sigma_N^{\rm SD} = 3 \lambda_{\rm SD}^2 \mu_N^2 / \pi$, and as in the previous case we can invert this equation to express the bound on $\lambda_{\rm SD}$ provided by Eq.~(\ref{sol1}) in terms of $\sigma_N^{\rm SD}$, where $N$ is either proton or neutron. Here we will use $\mathfrak{c}^N_4(m_\chi)$ as in \Eq{cfrakSD}, as well as the Mathematica interpolated functions $\mathcal{Y}_{4,4}^{(N,N')}(m_\chi)$. Explicitly, using
\beq
\lambda_\text{B}(m_\chi)^2 = \sigma_N^{\rm SD} \cdot \frac{256}{3} \pi m_\chi^2 \frac{m_N^2}{\mu_N^2} \mathcal{Y}_{4,4}^{(N,N)}(m_\chi)
\eeq
into ${\rm TS}(\lambda_{\rm B}, m_\chi)$ will provide the desired bound for the nucleon $N$, either proton or neutron.

\medskip

\begin{figure}[t]
\includegraphics[width=.315\textwidth]{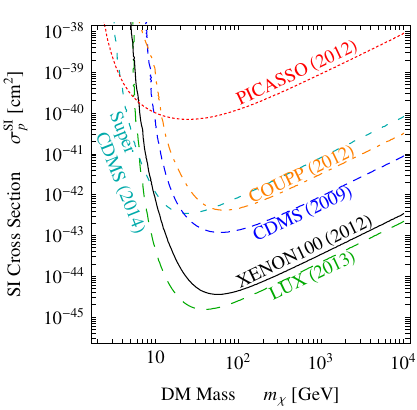}\quad
\includegraphics[width=.315\textwidth]{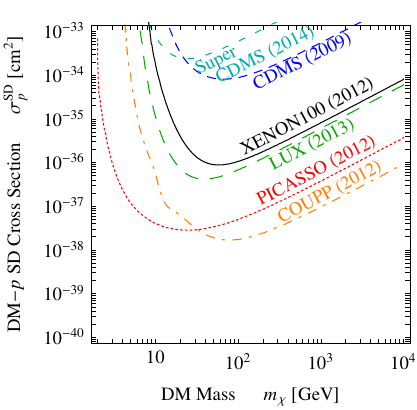}\quad
\includegraphics[width=.315\textwidth]{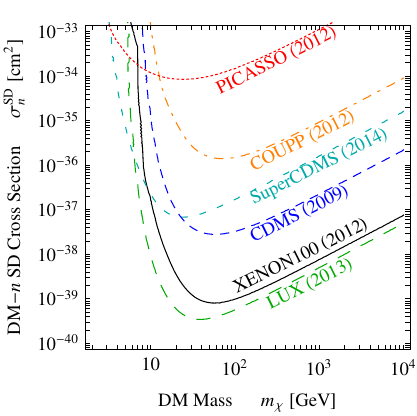}
\caption{\em \small \label{fig:standardbounds} 
The {\bfseries bounds on standard SI and standard SD cross sections} obtained within our formalism.}
\end{figure}

Our results for both SI and SD cross sections are plotted in \Fig{fig:standardbounds}, and they reproduce remarkably well the results given by the experimental collaborations in the references cited in Table~\ref{tableexp}. The only case where we find a notable difference of our bounds with the ones published by the experimental collaborations is for the spin-dependent interaction with protons in xenon. The reason for this difference has to be found in the uncertainty to which the xenon form factor for this interaction is known; in fact, as shown in Fig.~1 of \Ref{Aprile:2013doa}, results found by different collaborations disagree even by orders of magnitude.

While the agreement is good even in the small DM mass region ($m_\chi \lesssim 10$ GeV), we warn that that is the most critical one, where dedicated analyses have been performed by the experimental collaborations and by independent groups (see \eg \Ref{Akerib:2010pv, Ahmed:2010wy, Angle:2011th, Sorensen:2010hq, Davis:2012vy, Agnese:2013lua}).

\subsection{Minimal Dark Matter-inspired model}

As a final example, we consider a more involved model in which more than one operator contributes at the same time. 

The Minimal Dark Matter (MDM)~\cite{Cirelli:2005uq,Cirelli:2008id,Cirelli:2007xd,Cirelli:2009uv} construction consists in adding to the Standard Model (SM) the minimal amount of new physics (just one extra EW multiplet ${\cal X}$) and searching for the minimal assignments of its quantum numbers (spin, isospin and hypercharge) that make it a good Dark Matter candidate without ruining the positive features of the SM. 
The theory univocally selects a fermionic 5-plet of SU(2)$_{\rm L}$ with hypercharge $Y=0$ as the best candidate~\cite{Cirelli:2009uv}.

Of course, the `pure' MDM model as such is not of interest for our current analysis since it has no free parameters and therefore no parameter on which we can compute the bounds. We will therefore consider a modified version in which we assume that DM has a reduced SU(2) coupling $\tilde g = \epsilon \, g$, where $g$ is the ordinary SU(2)$_{\rm L}$ gauge coupling. We will thus be able to compute the bounds from different experiments on the quantity $\epsilon$. Moreover, while in the pure MDM model the mass of ${\cal X}$ is determined by the relic density requirement, here we leave it as a free parameter.

\medskip

\begin{figure}[t]
\includegraphics[width=0.31\textwidth]{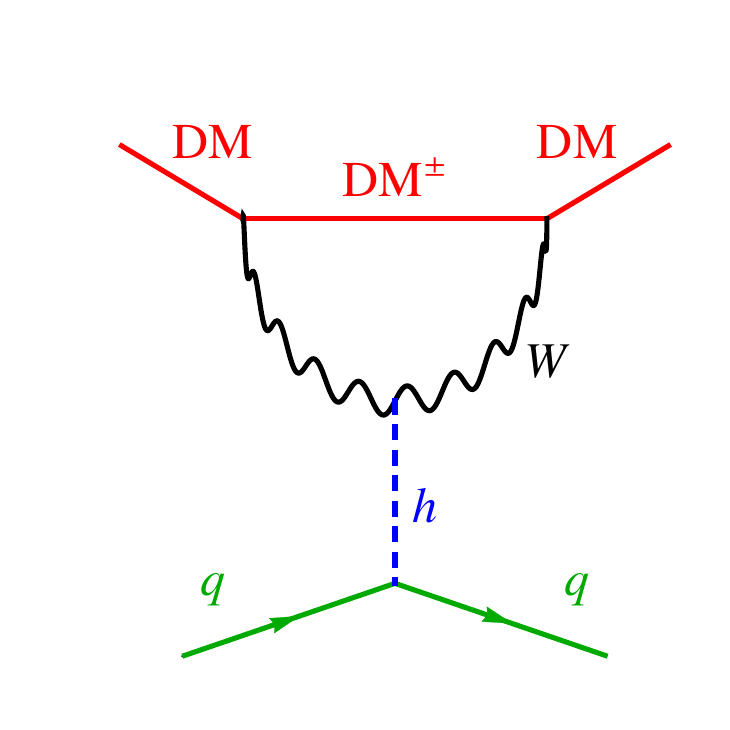} \quad
\includegraphics[width=0.31\textwidth]{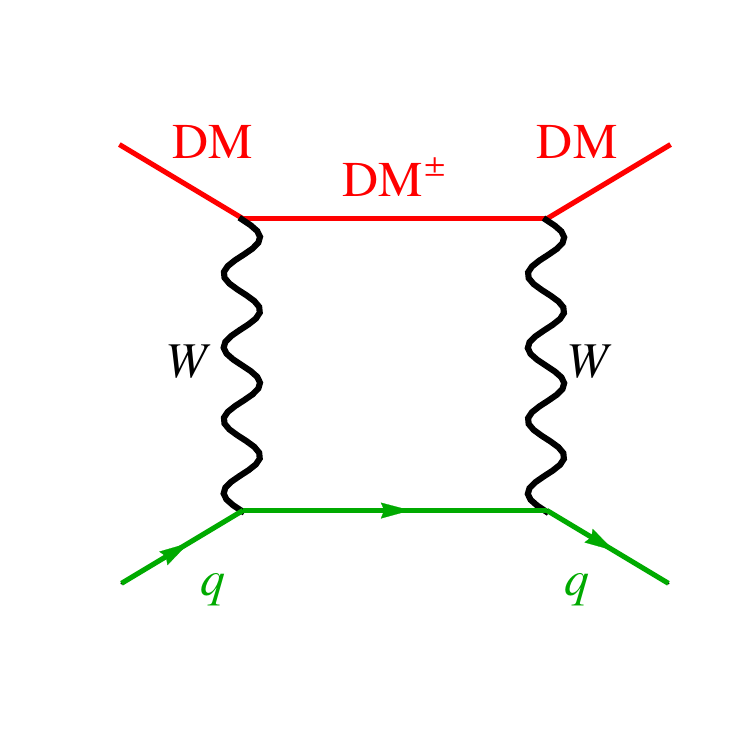} \quad
\includegraphics[width=0.31\textwidth]{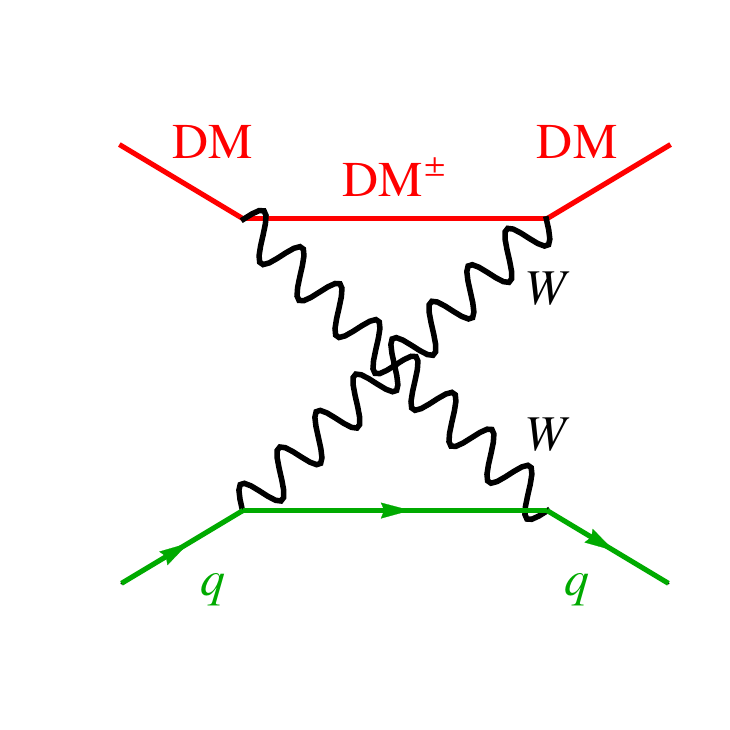}
\vspace{-1.2cm}
\caption{\label{fig:FeynLoop}\em {\bfseries One loop} DM-quark scattering for fermionic {\bfseries MDM} with $Y=0$. Figure adapted from~\cite{Cirelli:2005uq}.}
\end{figure}
\begin{figure}[t]
\includegraphics[width=0.31 \textwidth]{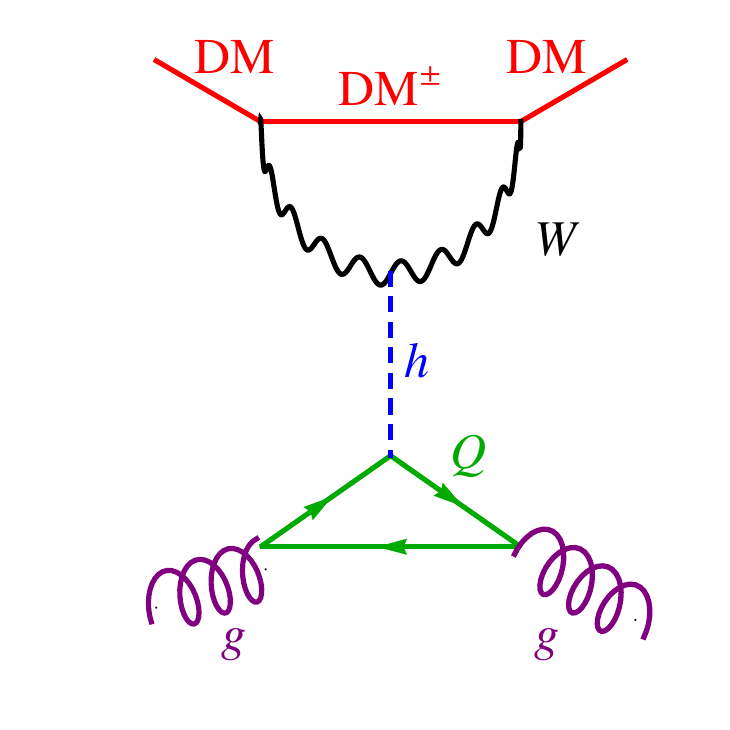} \quad
\includegraphics[width=0.31 \textwidth]{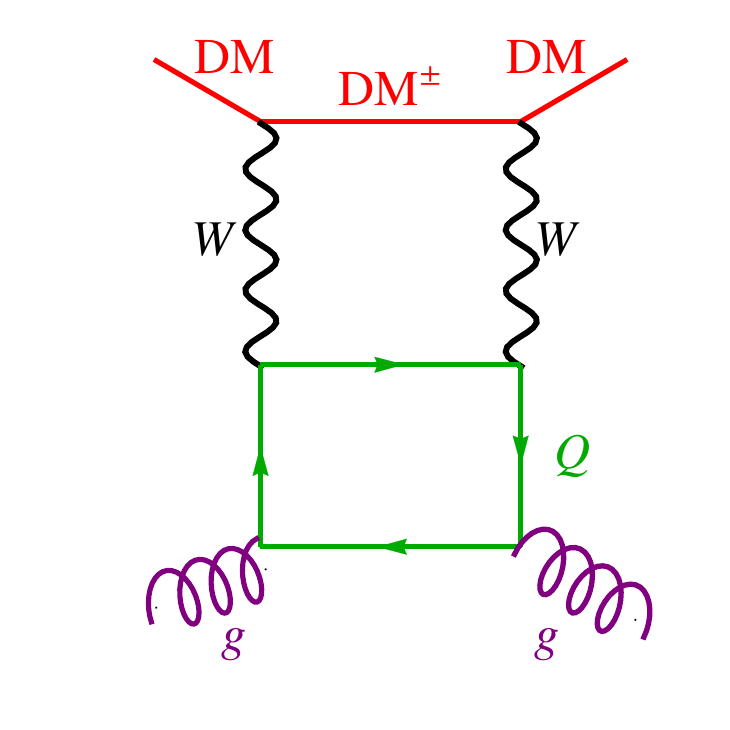} \quad
\includegraphics[width=0.31 \textwidth]{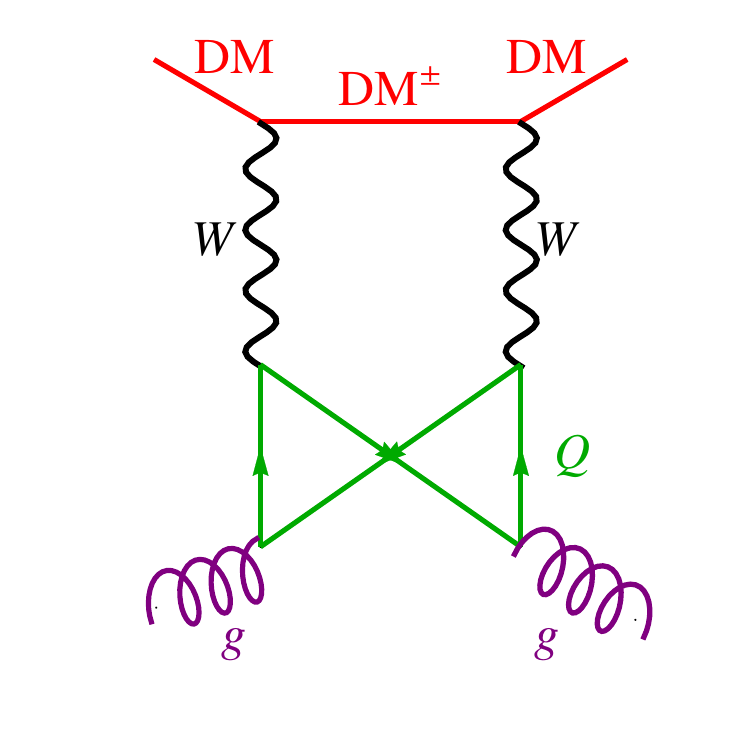}
\caption{\label{fig:Feyn2Loop}\em {\bfseries Two loop} DM-gluon scattering for fermionic {\bfseries MDM} with $Y=0$. Figure adapted from~\cite{Hisano:2011cs}.}
\end{figure}

\begin{figure}[t]
\begin{centering}
\includegraphics[width=0.5\textwidth]{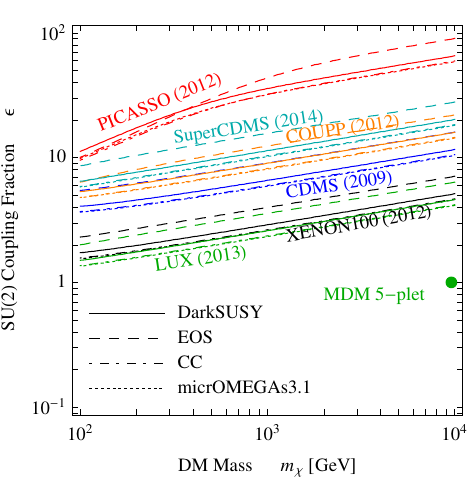}
\caption{\label{fig:MDMinspired}\em Bounds on $\epsilon$ in {\bfseries MDM-inspired model}. For each experiment we use four different sets of values of the scalar $f^{(N)}_{Tq}$, $f^{(N)}_{TG}$ and axial charges $\Delta^{(N)}_q$: from DarkSUSY \cite{Gondolo:2004sc}, EOS \cite{Ellis:2008hf}, CC \cite{Cheng:2012qr} and micrOMEGAs3.1 \cite{Belanger:2013oya} (see Table \ref{nuclearities} and Appendix \ref{qg2N}). The green dot marks the location of the `pure' MDM candidate with $\epsilon =1$ and mass fixed to $m_{\cal X} = 9.6$ TeV by the relic density requirement \cite{Cirelli:2009uv}.}
\end{centering}
\end{figure}

The MDM quintuplet scatters elastically on nuclei via the one loop diagrams in Fig.~\ref{fig:FeynLoop} and via the two loops diagrams in Fig.~\ref{fig:Feyn2Loop}. Upon a careful computation\footnote{These results are based on the computations in \Ref{Cirelli:2005uq} and~\cite{Hisano:2011cs} (for the one loop diagrams) and in \Ref{Hisano:2011cs} (for the two loops diagrams). We always work in the limit of $M_W, m_t \ll m_{\cal X}$, where $M_W$ is the $W$ mass and $m_t$ is the top mass, relevant for the two loop diagrams. One can check that the term containing the twist-2 operator gives rise to the term proportional to $1/M_W^3$ in the original computation of~\cite{Cirelli:2005uq}, however with an opposite sign and with an $o(1)$ difference in the coefficient of which we are not able to pin down the origin: we adopt the more recent result of~\cite{Hisano:2011cs}). The different sign causes a cancellation between the term in $1/M_Wm_h^2$ and the term in $1/M_W^3$, which significantly suppresses the cross section~\cite{Hisano:2011cs}. Also, for the coefficient of the $\Op^q_8$ operator we have a discrepancy of a factor of 2 with the corresponding computation in~\cite{Hisano:2011cs} of which we are not able to pin down the origin: we stick to the result in~\cite{Cirelli:2005uq}. Finally, in principle one should also consider the twist-2 gluon operator $G^a_{\mu\rho} G^a_{\rho\nu} + (g_{\mu\nu} / 4) G^a_{\alpha\beta} G^a_{\alpha\beta}$. This has been done \eg in Ref.~\cite{Hill:2011be}. Due to the aforementioned cancellation between terms in the scattering amplitude, its contribution might be relevant while dominated by uncertainties. Our estimate shows it to be subdominant anyway with respect to the gluon operator that we include: following \cite{Hisano:2012wm}, we do not include it in this analysis.} one arrives at determining the high-energy effective scattering Lagrangian 
\begin{equation}
\label{MDMlagrangian}
\begin{aligned}
\Lag_{\rm eff}=\, & \frac{3}{32\, \pi} \frac{\epsilon^2 \, g^4}{M_W}\sum_q
\left[
\frac{1}{m_h^2} \, m_q \, \bar {\cal X}{\cal X} \ \bar{q}{q} - \frac{2}{3 m_{\cal X}} \,
{ \bar {\cal X} \gamma^\mu\gamma^5 {\cal X} \ \bar{q} \gamma_\mu\gamma^5 {q} } \right] \\
& -\frac{1}{16 \, \pi} \frac{\epsilon^2 \, g^4}{M_W^3\, m_{\cal X}} \, \bar {\cal X}(i\partial^\mu\gamma^\nu){\cal X} \left[ \frac{1}{2} \, \bar q \, i (D_\mu \gamma_\nu + D_\nu \gamma_\mu -\frac{1}{2} g_{\mu\nu} \slashed{D})q \right] \\
& - \frac{1}{16\, \pi}\frac{\epsilon^2 \, g^4}{M_W}\left( \frac{\sum_Q \kappa_Q}{m_h^2} +\frac{1}{M_W^2} \right)\frac{\alpha_s}{8\, \pi} \, \bar {\cal X}  {\cal X} \ G^a_{\mu\nu} G^a_{\mu\nu} \ ,
\end{aligned}
\end{equation}
where $Q=c,b,t$ in the last line, with $\kappa_c=1.32, \kappa_b=1.19, \kappa_t=1$~\cite{Hisano:2011cs}.
As apparent, the effective Lagrangian contains three dominant operators among those listed in Eq.~(\ref{DOperators})-(\ref{DOperatorsg}): $\Op^q_1$, $\Op^q_8$ and $\Op^g_1$. 
It also contains the term with a twist-2 quark operator (let us denote the full operator as $\Op^q_A = \bar {\cal X}(i\partial^\mu\gamma^\nu){\cal X} \, \Op^{q,\text{twist-2}}_{\mu\nu}$), which is not among the standard ones listed above and therefore has to be reduced to its NR limit explicitly.
The coefficients explicitly read
\begin{subequations}
\begin{align}
\label{MDMcRel}
c_1^q &= \epsilon^2 \frac{3}{32 \, \pi} \frac{g^4}{M_W} m_q \frac{1}{m_h^2} \ , \\ 
c_8^q &= -\epsilon^2 \frac{1}{16 \, \pi} \frac{g^4}{M_W m_{\cal X}} \ , \\
c_{A}^q &= -\epsilon^2 \frac{1}{16 \, \pi} \frac{g^4}{M_W^3 m_{\cal X}} \ , \\
c_1^g &=  \epsilon^2 \frac{1}{16\, \pi} \frac{g^4}{M_W} \left( \frac{3.51}{m_h^2} +\frac{1}{M_W^2} \right) \ .
\end{align}
\end{subequations}
$\Op^q_A$ contributes $\Op^N_1$~\cite{Hisano:2011cs} and therefore, at the nucleon level, the effective operators $\Op^N_1$ and $\Op^N_8$ have coefficients (see Eq.~(\ref{c^N_k}))
\begin{subequations}
\begin{align}
\label{MDMcN}
c_1^N &= \epsilon^2 \frac{3}{32 \, \pi} \frac{g^4}{M_W} m_N \left[  \frac{1}{m_h^2} \left( \sum_{q=u,d,s} f^{(N)}_{Tq} +\frac{2}{27} f^{(N)}_{TG} (1 - \tfrac{2}{3} {\textstyle \sum_Q} \kappa_Q) \right) \right. \nonumber \\
 & \left. \phantom{= \epsilon^2 \frac{3}{32 \, \pi} \frac{g^4}{M_W} m_N \ \ } -\frac{1}{M^2_W} \left( \frac{4}{81} f^{(N)}_{TG}+\frac{1}{2} \sum_{q=u,d,s,c,b} (q^{{(N)}}(2)+\bar q^{{(N)}}(2)) \right)  \right] \ , \\
c_8^N &= -\epsilon^2 \frac{1}{16 \, \pi}  \frac{g^4}{M_W m_{\cal X}} \sum_{q} \Delta^{(N)}_q \ . 
\end{align}
\end{subequations}
Here $q^{{(N)}}(2)$ and $\bar q^{{(N)}}(2)$ are the second moment of the parton distribution functions for a quark $q$ or antiquark $\bar q$ in the nucleon $N$, whose values are given \eg in~\cite{Hisano:2011cs}. For the proton one has $u^{{(p)}}(2) = 0.22, d^{{(p)}}(2) = 0.11, \bar u^{{(p)}}(2) = 0.034, \bar d^{{(p)}}(2) = 0.036, s^{{(p)}}(2) = \bar s^{{(p)}}(2) = 0.026, c^{{(p)}}(2) =\bar c^{{(p)}}(2) = 0.019, b^{{(p)}}(2) =\bar b^{{(p)}}(2) = 0.012$ (for the neutron one needs to exchange the values for the up quark with those for the down quark).
Making use of Eq.~(\ref{fromNtoNR}), one determines that the NR operators for the model are $\Op^{\rm NR}_1$ and $\Op^{\rm NR}_4$, with coefficients
\begin{subequations}
\begin{align}
\mathfrak{c}_1^N (\epsilon, m_{\cal X}) &= \epsilon^2 \frac{3}{8 \, \pi} \frac{g^4}{M_W} m^2_N m_{\cal X} \left[  \frac{1}{m_h^2} \left( \sum_{q=u,d,s} f^{(N)}_{Tq} +\frac{2}{27} f^{(N)}_{TG} (1 - \tfrac{2}{3} {\textstyle \sum_Q} \kappa_Q) \right) \right. \nonumber \\
 & \left. \phantom{= \epsilon^2 \frac{3}{32 \, \pi} \frac{g^4}{M_W} m_N \ \ } -\frac{1}{M^2_W} \left( \frac{4}{81} f^{(N)}_{TG}+\frac{1}{2} \sum_{q=u,d,s,c,b} (q^{{(N)}}(2)+\bar q^{{(N)}}(2)) \right)  \right] \ , \\
\mathfrak{c}_4^N (\epsilon, m_{\cal X}) &= \epsilon^2  \frac{g^4}{\pi} \frac{m_N}{M_W} \sum_{q} \Delta^{(N)}_q \ ,
\end{align}
\end{subequations}
where the notation makes apparent that the role of the generic variable $\lambda$ (employed in Sec.~\ref{sec:formfactors}) is here played by the $\epsilon$ parameter. At this point we can make explicit use of Eq.~(\ref{sol1}) and write an expression for $\lambda_{\rm B}^2$ involving the above coefficients and the ${\cal Y}_{1,1}^{(N,N')}(m_{\cal X})$ and ${\cal Y}_{4,4}^{(N,N')}(m_{\cal X})$ rescaling functions provided in \href{http://www.marcocirelli.net/NRopsDD.html}{the website}. We just plug this explicit expression for $\lambda^2_{\rm B}$ into the ${\rm TS}$ so that we have a ${\rm TS}(\epsilon, m_{\cal X})$ and draw a contour plot in order to get the resulting bound on $\epsilon$. In Fig.~\ref{fig:MDMinspired} we show the result. This figure illustrates, for instance, that fairly large values of $\epsilon$ are still allowed by the current experiments.\footnote{Or that, in turn, the pure MDM case is still far from the experimental sensitivity. This conclusion is different from that in~\cite{Cirelli:2005uq} because of the reduction in the cross section due to the cancellations discussed above.}

\section{Conclusions}
\label{sec:conclusions}

In this paper we have described a method and provided a self consistent set of tools for obtaining bounds from some of the current leading direct detection experiments on virtually any DM model. The method is based on the use of non-relativistic operators to describe the interaction between DM and ordinary matter. It builds on the results obtained by~\Ref{Fitzpatrick:2012ix} and incorporates into them the necessary detector and astrophysical ingredients. Indeed, our main result consists in building up a set of {\em integrated form factors} $\tilde{\mathcal{F}}$ which allow to compute the detector response for any combination of NR operators. Our main output, in turn, consists in providing (on \href{http://www.marcocirelli.net/NRopsDD.html}{this website}) a set of Test Statistic functions (one for each experiment we consider) and a set of functions $\mathcal{Y}$, which are just a rescaling of the integrated form factors. Armed with these, one can obtain a constraint on any parameter of a DM model by using the recipe spelled out in Sec.~\ref{sec:examples}. 
The TS functions that we provide for each experiment reflect the current status of results: they will change whenever new results are released and we will update them accordingly. On the other hand, the integrated form factor (and, a fortiori, the scaling functions) change only if major changes occur in the experimental set-ups, \eg in efficiencies, cuts, and thresholds. We do not foresee frequent updates for those, but we will consider them when necessary.
\\
As the NR operators describe any possible DM interaction with ordinary matter, the method is fully model-independent and can be applied to any DM high energy construction. Indeed, somewhat as an aside, we also reviewed the procedure to pass from high-energy effective operators (into which a large class of models can be cast) to NR operators.

\medskip

We now conclude with a brief discussion of some features of the method and some future perspectives. We stress that the method is exact for `counting experiments', \ie those which can be considered as having a single energy bin (like \XENON\ and \CDMS), but the approximation for experiments with more bins is very well verified (see the discussion in Sec.~\ref{sec:constraints}). We remind that we considered only `null result' experiments, and only four of them, but clearly the method can be extended to any experiment. Also, the form factors for other nuclei potentially of interest for DD should be computed. Moreover, different statistical tools from the ones we employed could be used.
We encourage the experimental collaborations to release, when possible, their own likelihood and to present their computation of the integrated form factors, enriched by a knowledge of their detector deeper than that available to us. 

\medskip

Striving to devise model-independent methods for DM phenomenology (in direct detection but also more generally) is a crucial direction to pursue in order to make sense of the current exciting experimental panorama.

{\small 
\paragraph{Acknowledgments.}

\noindent E.D.N.~heartily thanks Liam Fitzpatrick and Ami Katz for useful discussions and explanations about their works \cite{Fitzpatrick:2012ix, Fitzpatrick:2012ib}. We also thank Vincenzo Cirigliano and Andrea De Simone for discussions, and Giorgio Busoni for checking our code and pointing out two problems with three plots in \Fig{fig:boundsindividual} (now fixed). This work is supported by the European Research Council ({\sc Erc}) under the EU Seventh Framework Programme (FP7/2007-2013) / {\sc Erc} Starting Grant (agreement n. 278234 - `{\sc NewDark}' project). M.C.~is supported in part by the French national research agency {\sc Anr} under contract {\sc Anr} 2010 {\sc Blanc} 041301 and by the EU ITN network {\sc Unilhc}. E.D.N.~is supported in part by DOE grant DE-FG02-13ER42022. M.C. acknowledges the hospitality of the Institut d'Astrophysique de Paris ({\sc Iap}) and the Theory Division of CERN where a part of this work was done. E.D.N.~wishes also to thank M.C.~and P.P.. But M.C.~wishes to thank E.D.N.~and P.P.. In turn P.P., guess what, thanks E.D.N.~and M.C..}

\appendix

\section{Velocity integral}
\label{sec:velocityntegral}

The DM velocity distribution $f_\text{E}(\vec v)$ in the Earth frame is related to the velocity distribution in the galactic frame $f_\text{G}(\vec v \, ')$ by the Galilean velocity transformation $f_\text{E}(\vec v) = f_\text{G}(\vec v + \vec v_\text{E})$, where $\vec v_\text{E}$ is the Earth velocity with respect to the galactic frame. One has that $\vec v_\text{E} = \vec v_\odot + \vec v_\oplus$, where $\vec v_\oplus$ is the Earth rotational velocity, and $\vec v_\odot$ the sum of the galactic rotational velocity of our local system and the Sun's proper motion. More details can be found for instance in Ref.~\cite{Fornengo:2003fm, Savage:2008er}.


In this paper we consider for $f_\text{G}(\vec v)$ a truncated Maxwell-Boltzmann, corresponding to an isothermal sphere density profile for the DM:
\beq
f_\text{G}({\vec v}) = \frac{\exp(- v^2 / v_0^2)}{(v_0 \sqrt{\pi})^3 \, \text{erf}(v_\text{esc} / v_0) - 2 v_0^3 \pi (v_\text{esc} / v_0) \exp(- v_\text{esc}^2 / v_0^2)} \ .
\rule[-12pt]{0pt}{20pt}
\eeq
The denominator is responsible for the normalization of the velocity distribution,
\beq
\int_{v \leqslant v_{\rm esc}} \ud^3 v \, f_\text{G}(\vec v) = 1 \ ,
\eeq
where $v_{\rm esc}$ denotes the escape velocity of DM particles from our galaxy. For definiteness, in this work we assume $v_{\rm esc} = 544$ km/s \cite{Smith:2006ym}.

At zeroth order in the non-relativistic expansion, the scattering amplitude does not depend on the DM velocity, and therefore from \Eq{sigmaT} one has $\ud \sigma_T / \ud \ER \propto v^{-2}$. This means that, when computing the rate in \Eq{RT}, one has to compute the integral
\beq\label{I0}
{\mathcal I_0(\vmin)} = \int_{v \geqslant v_{\rm min}(E_R)} \hspace{-.85cm} \ud^3 v \,\, \frac{f_\text{E}(\vec v)}v \ .
\eeq
When going at first order, the differential cross section might feature also a $v$-independent term, thus giving rise to
\beq\label{I1}
{\mathcal I_1(\vmin)} = \int_{v \geqslant v_{\rm min}(E_R)} \hspace{-.85cm} \ud^3 v \,\,v \, f_\text{E}(\vec v) \ .
\eeq
For a truncated Maxwell-Boltzmann velocity distribution, these integrals can be computed analytically and result in \cite{Barger:2010gv}
\beq\label{velocityI}
{\mathcal I_0(\vmin)}
=
\frac1{2\,v_0 \eta_\text{E}}\left[\mbox{erf}(\eta_+)-\mbox{erf}(\eta_-)\right] -\frac1{\sqrt\pi \,v_0 \eta_\text{E}}\left(\eta_+-\eta_-\right)e^{-\eta_{\rm esc}^2}
\eeq
and
\begin{gather}
{\mathcal I_1(\vmin)}
=
v_0\left[\left(\frac{\eta_-}{2\sqrt\pi\,\eta_\text{E}}+\frac1{\sqrt\pi}\right)e^{-\eta_-^2}-\left(\frac{\eta_+}{2\sqrt\pi\,\eta_\text{E}}-\frac1{\sqrt\pi}\right)e^{-\eta_+^2}\right] \nonumber
\\
\label{velocityI1} 
+\frac{v_0}{4\,\eta_\text{E}}\left(1+2\eta_\text{E}^2\right)\left[\mbox{erf}(\eta_+)-\mbox{erf}(\eta_-)\right] \rule{0pt}{20pt}
\\
-\frac{v_0}{\sqrt\pi}\left[2+\frac1{3\eta_\text{E}} \left(\left(\eta_{\rm min}+\eta_{\rm esc}-\eta_-\right)^3-\left(\eta_{\rm min}+\eta_{\rm esc}-\eta_+\right)^3\right)\right]e^{-\eta_{\rm esc}^2} \ , \nonumber
\end{gather}
where we defined the normalized velocities
\begin{gather}
\eta_\text{E} \equiv v_\text{E} / v_0 \ ,
\\
\eta_{\rm esc} \equiv v_{\rm esc} / v_0 \ ,
\\
\eta_{\rm min}(E_R) \equiv v_{\rm min}(E_R) / v_0 \ ,
\end{gather}
and
\beq
\eta_\pm(E_R) = \min (\eta_{\rm min}(E_R) \pm \eta_\text{E}, \eta_{\rm esc}) \ .
\eeq

\section{From quarks and gluons to nucleons}\label{qg2N}
Given the very low energies involved in DM scattering with matter, the relevant degrees of freedom to be considered are not quarks and gluons, but rather nucleons and nuclei. In \Sec{sec:formfactors} we recalled the formalism introduced in \Ref{Fitzpatrick:2012ix}, allowing one to write the DM-nucleus cross section from the interaction with the nucleons. However, many models are formulated in terms of interactions with the fundamental degrees of freedom, namely quarks and gluons. In this Appendix we summarize the formulas connecting amplitudes at the quark and gluon level with the matrix element with the nucleons, and provide connections to the relevant literature. In particular, we will review the gluon matrix elements $\langle N | G^a_{\mu\nu} G^a_{\mu\nu} | N \rangle$ and $\langle N | G^a_{\mu\nu} \tilde{G}^a_{\mu\nu} | N \rangle$, and the matrix elements of quark bilinears $\langle N | \bar q \Gamma q | N \rangle$ with $\Gamma$ either $\uno, \gamma^5, \gamma^\mu, \gamma^\mu \gamma^5, \sigma^{\mu\nu}$ or $\sigma^{\mu\nu} \gamma^5$. Whereas these quantities can be extracted from experimental data, an effort has been recently undertaken to compute them within lattice simulations. However, still no general agreement has been reached in some cases, and quantities such as the strange quark contribution to the nucleon mass and spin are still plagued by sizeable uncertainties. For a comparison, we collect some of the numerical values reported by few different collaborations in Table \ref{nuclearities}. We consider four standard references, Ref.~\cite{Ellis:2000ds, Gondolo:2004sc, Ellis:2008hf, Belanger:2008sj}, and the two more recent \Ref{Cheng:2012qr, Belanger:2013oya} (notice that \Ref{Belanger:2013oya} is the \virg{update} of \cite{Belanger:2008sj}).

\begin{table}
\centering
\begin{tabular}{c|c|c|c|c|c|c|c|c}
Ref. & $f_{Tu}^{(p)}$ & $f_{Tu}^{(n)}$ & $f_{Td}^{(p)}$ & $f_{Td}^{(n)}$ & $f_{Ts}^{(N)}$ & $\Delta_u^{(p)}$ & $\Delta_d^{(p)}$ & $\Delta_s^{(p)}$ \\
\hline
\cite{Gondolo:2004sc} & $0.023$ & $0.019$ & $0.034$ & $0.041$ & $0.14$ & $0.77$ & $-0.40$ & $-0.12$ \\
\cite{Ellis:2008hf} & $0.027$ & $0.022$ & $0.039$ & $0.049$ & $0.36$ & $0.84$ & $-0.43$ & $-0.09$ \\
\cite{Cheng:2012qr} & $0.017$ & $0.012$ & $0.023$ & $0.033$ & $0.053$ & $0.84$ & $-0.44$ & $-0.03$ \\
\cite{Belanger:2013oya} & $0.015$ & $0.011$ & $0.019$ & $0.027$ & $0.045$ & $0.84$ & $-0.43$ & $-0.085$ \\
\cite{Ellis:2000ds} & $0.020$ & $0.014$ & $0.026$ & $0.036$ & $0.118$ & $0.78$ & $-0.48$ & $-0.15$ \\
\cite{Belanger:2008sj} & $0.023$ & $0.018$ & $0.033$ & $0.042$ & $0.26$ & $0.84$ & $-0.43$ & $-0.08$ \\
\end{tabular}
\caption{\em \small \label{nuclearities} A compilation of values of the {\bfseries scalar and axial charges} from the indicated references.}
\end{table}

\subsection{Scalar couplings}
The matrix element of the scalar quark bilinear $\langle N | \bar q q | N \rangle$, at zero momentum transfer, can be computed in the following way \cite{Shifman:1978zn}. At zero momentum $m_N = \NN{\Theta_{\mu\mu}}$, where
\beq\label{thetamumu}
\Theta_{\mu\mu} = \sum_q m_q \, \bar q q + \frac{\beta(\alpha_s)}{4 \alpha_s} G^a_{\mu\nu} G^a_{\mu\nu}
\eeq
is the trace of the QCD energy-momentum tensor after applying the equations of motion; the gluon contribution comes from the trace anomaly. Now one can integrate out the heavy quarks $h = c, b, t$ via the heavy quarks expansion, yielding at the lowest order a result that is reproduced by the substitution
\beq\label{heavyquarks}
m_h \, \bar h h \rightarrow - \frac{\alpha_s}{12 \pi} G^a_{\mu\nu} G^a_{\mu\nu} \ .
\eeq
Expanding the beta function in powers of $\alpha_s$ we get finally, at the lowest order,
\beq
\Theta_{\mu\mu} = \sum_{q = u, d, s} m_q \, \bar q q - \frac{9 \alpha_s}{8 \pi} G^a_{\mu\nu} G^a_{\mu\nu} \ .
\eeq
The gluon contribution can then be expressed in terms of light quarks via
\beq\label{gluonstolightq}
- \frac{1}{m_N} \frac{9 \alpha_s}{8 \pi} \NN{G^a_{\mu\nu} G^a_{\mu\nu}} = 1 - \sum_{q = u, d, s} f_{Tq}^{(N)} \equiv f_{TG}^{(N)} \ ,
\eeq
where the quantities
\beq
f_{Tq}^{(N)} \equiv \frac{\langle N | m_q \, \bar q q | N \rangle}{m_N}
\eeq
express the light quark contributions to the nucleon mass. Values of the $f_{Tq}^{(N)}$ coefficients reported by a few different collaborations are collected in Table \ref{nuclearities}.

We are now interested in the coupling of a scalar operator $S$ to the nucleon, its interaction with the quarks being dictated by the Lagrangian
\beq
\Lag_{Sq} = S \sum_q c_q \, \bar q q
\eeq
with $c_q$ the interaction coefficients. This interaction induces a coupling of $S$ to gluons via a quark loop, so to be as generic as possible we can also add an independent $S$-gluons scalar interaction with the same form \cite{Shifman:1978zn},
\beq
\Lag_{Sg} = \frac{c_g}{\Lambda} \frac{\alpha_s}{12 \pi} \, S \, G^a_{\mu\nu} G^a_{\mu\nu} \ ,
\eeq
where the numerical factors have been chosen for later convenience; this coupling is generated again at the loop level, $\Lambda$ being connected to the mass of the particles running in the loop, and $c_g$ being some coefficient whose value reflects the underlying process. The $S$-nucleon interaction can be written now as
\beq
\Lag_{SN} = \NN{\Lag_{Sq} + \Lag_{Sg}} \bar{N} N \equiv c_N \, S \, \bar{N} N \ ,
\eeq
with $N$ the nucleon field, and where using Eq.~\eqref{heavyquarks} and \eqref{gluonstolightq} we can express $c_N$ as
\beq\label{DDscalarcoupling-masterformula}
c_N = \sum_{q = u, d, s} c_q \frac{m_N}{m_q} f_{Tq}^{(N)} + \frac{2}{27} f_{TG}^{(N)} \left( \sum_{q = c, b, t} c_q \frac{m_N}{m_q} - c_g \frac{m_N}{\Lambda} \right) \ .
\eeq

\subsection{Pseudoscalar couplings}
\label{sec:pseudoscalar}
The pseudoscalar coupling of an operator $P$ with quarks is
\beq
\Lag_{Pq} = P \sum_q c_q \, \bar q \, i \gamma^5 q \ .
\eeq
Integrating out the heavy quarks $h = c, b, t$, one has a loop-induced coupling of $P$ with gluons as in the previous case, reproduced by the substitution valid at lowest order \cite{Shifman:1978zn}
\beq
m_h \, \bar h \, i \gamma^5 h \rightarrow - \frac{\alpha_s}{16 \pi} G^a_{\mu\nu} \tilde{G}^a_{\mu\nu} \ ,
\eeq
where $\tilde{G}^a_{\mu\nu} \equiv \varepsilon_{\mu\nu\rho\sigma} G^a_{\rho\sigma}$. We can also add, as above, an independent $P$-gluons coupling, generated for instance by other particles in the loop. This can be written effectively as
\beq
\Lag_{Pg} = \frac{c_g}{\Lambda} \frac{\alpha_s}{8 \pi} \, P \, G^a_{\mu\nu} \tilde{G}^a_{\mu\nu} \ ,
\eeq
where again $\Lambda$ is connected to the mass of the particles running in the loop, and the numerical factors have been chosen for later convenience.

Evaluating the gluonic operator between nucleon states is a problematic task; we rely on the analysis performed in \cite{Cheng:1988im, Cheng:2012qr}, based on the relation
\beq
\langle N | \bar{u} \, i \gamma^5 u + \bar{d} \, i \gamma^5 d + \bar{s} \, i \gamma^5 s |N \rangle = 0 \ ,
\eeq
derived from the large-$N_{\rm c}$ and chiral limits. This leads to
\beq
\left< N \left| \frac{\alpha_s}{8 \pi} G^a_{\mu\nu} \tilde{G}^a_{\mu\nu} \right| N \right> = m_N \bar{m} \sum_{q = u, d, s} \frac{\Delta_q^{(N)}}{m_q} \ ,
\eeq
where $\bar{m} \equiv (1 / m_u + 1 / m_d + 1 / m_s)^{-1}$. The coefficients $\Delta_q^{(N)}$, defined by $2 \Delta_q^{(N)} s^\mu = \langle N | \bar{q} \gamma^\mu \gamma^5 q | N \rangle$ with $s^\mu$ the nucleon spin four-vector, parametrize the quark spin content of the nucleon $N$. They involve therefore an axial-vector quark operator rather than a pseudoscalar one, but the two are related by PCAC \cite{Cheng:1988im}. These coefficients are argued to be negligible for heavy quarks \cite{Polyakov:1998rb}, while for light quarks they satisfy the following relations: $\Delta_u^{(p)} = \Delta_d^{(n)}$, $\Delta_d^{(p)} = \Delta_u^{(n)}$, $\Delta_s^{(p)} = \Delta_s^{(n)}$. As for the scalar charges $f_{Tq}^{(N)}$, we collect in Table \ref{nuclearities} values of the coefficients $\Delta_q^{(N)}$ reported by few different collaborations.

The $P$-nucleon interaction can be now written as
\beq
\Lag_{PN} = \NN{\Lag_{Pq} + \Lag_{Pg}} \bar{N} i \gamma^5 N \equiv c_N \, P \, \bar{N} i \gamma^5 N \ ,
\eeq
yielding \cite{Cheng:2012qr}
\beq
c_N = \sum_{q = u, d, s} \frac{m_N}{m_q} \left[ (c_q - C) + c_g \frac{\bar{m}}{\Lambda} \right] \Delta_q^{(N)} \ ,
\eeq
where we defined $C \equiv \sum_q c_q \, \bar{m} / m_q$.

\subsection{Vector couplings}
The coupling of a vector operator $V_\mu$ with the quark current $\bar{q} \gamma^\mu q$,
\beq
\Lag_{Vq} = c_q \, V_\mu \, \bar{q} \gamma^\mu q \ ,
\eeq
leads to the same vector interaction with the nucleon current,
\beq
\Lag_{VN} = c_N \, V_\mu \, \bar{N} \gamma^\mu N \ .
\eeq
Since the quark current is conserved, the nucleon charge is obtained by the quark charges by merely summing over the valence quarks of the nucleons. The interaction coefficients are therefore $c_p = 2 c_u + c_d$ and $c_n = c_u + 2 c_d$, for proton and neutron respectively.

\subsection{Axial-vector couplings}\label{A-V}
An axial-vector interaction of quarks with an operator $A_\mu$,
\beq
\Lag_{Aq} = c_q \, A_\mu \, \bar{q} \gamma^\mu \gamma^5 q \ ,
\eeq
leads to an axial-vector interaction with nucleons
\beq
\Lag_{AN} = c_N \, A_\mu \, \bar{N} \gamma^\mu \gamma^5 N \ ,
\eeq
with $c_N = \sum_q c_q \Delta_q^{(N)}$. The quantities $\Delta_q^{(N)}$ have been introduced in \Sec{sec:pseudoscalar}, and values quoted by a few collaborations can be found in Table \ref{nuclearities}.

\subsection{Tensor couplings}
The quark tensor interaction with an operator $T_{\mu\nu}$,
\beq
\Lag_{Tq} = c_q \, T_{\mu\nu} \, \bar{q} \, \sigma^{\mu\nu} q \ ,
\eeq
leads to a nucleon tensor interaction
\beq
\Lag_{TN} = c_N \, T_{\mu\nu} \, \bar{N} \, \sigma^{\mu\nu} N \ ,
\eeq
with $c_N = \sum_q c_q \delta_q^{(N)}$ \cite{He:1994gz}. Using the relation $\sigma^{\mu\nu} \gamma^5 = (i / 2) \epsilon^{\mu\nu\rho\tau} \sigma_{\rho\tau}$ one can also find the relative expression for an axial-tensor coupling. The tensor charges $\delta_q^{(N)}$ are interpreted as the difference between the spin of quarks and the spin of anti-quarks in nucleons. They have been recently measured by \Ref{Anselmino:2008jk} as $\delta_u^{(p)} = 0.54$, $\delta_d^{(p)} = -0.23$ at $Q^2 = 0.8$ GeV$^2$; a newer global analysis with more data has been presented by the same collaboration in \Ref{Anselmino:2013vqa}, with results compatible with the previous ones within the quoted uncertainties. \Ref{Belanger:2008sj, Belanger:2013oya} quote instead the values $\delta_u^{(p)} = 0.84$, $\delta_d^{(p)} = -0.23$, and $\delta_s^{(p)} = -0.05$. See also~\cite{Bacchetta:2012ty} for other determinations of these values. Supposedly, the proton tensor charges are related to the neutron ones in the same way as the axial charges $\Delta_q^{(N)}$ are.

\newpage\normalsize

\phantomsection
\addcontentsline{toc}{section}{Addenda:}

\renewcommand*{\thesection}{\arabic{section}}

\setcounter{section}{0}
\renewcommand*{\theHsection}{\the\value{section}}

\section{Addendum: LUX 2013 data}
\label{LUX}

On October 30$^{\rm th}$, 2013, the \LUX\ collaboration announced their first DM search results in \Ref{Akerib:2013tjd}. With the data collected in about $85$ live-days, they were able to set a stronger bound on the spin-independent interaction cross section, with respect to the preexisting limits. Given the relevance of this result, we include it in our \href{http://www.marcocirelli.net/NRopsDD.html}{set of numerical tools} (Release 2) to derive bounds from direct DM searches. 

\subsection*{Description of \LUX}
\addcontentsline{toc}{subsection}{Description of \LUX}

The Large Underground Xenon (\LUX) experiment, operated at the Sanford Underground Research Facility in South Dakota, is a dual-phase xenon time-projection chamber. As the \XENON\ experiment, it uses the prompt scintillation ($S_1$) and ionization ($S_2$) signals to reconstruct the deposited energy and to discriminate nuclear recoils from electron recoils.

In \Ref{Akerib:2013tjd}, a non-blind analysis was conducted on data collected with an exposure $w$ of $85.3$ live-days $\times$ $118.3$ kg of fiducial volume. After cuts, $160$ events were found within the $S_1$ energy region $2 - 30$ photoelectrons. The collaboration found that all the events are compatible with the expected electron recoil background distribution.

Due to lack of detailed information on expected background and event distribution in the $S_1$--$S_2$ space, it is very difficult to devise a coherent and spectrum independent analysis of the data. In \Ref{DelNobile:2013gba}, bounds were computed by assuming that either $0$, $1$, $3$, $5$ or all the $24$ events below the electron recoil band in Fig.~4 of \Ref{Akerib:2013tjd} are indistinguishable from the expected signal, while the remaining are background events. The statistical analysis adopted in \Ref{DelNobile:2013gba} does not need information on the expected background, however our analysis (described in Sec.~\ref{sec:constraints}) requires that information and therefore we can not use the same approach. We proceed therefore as follows. Motivated by the fact that calibration neutrons are expected to mimic the $S_1$--$S_2$ signal of heavy DM particles in the detector, we assume that DM events on the $S_1$--$\log_{10}(S_2/S_1)$ plane distribute evenly above and below the mean of the nuclear recoil event distribution (solid red line in Fig.~3 and 4 of \Ref{Akerib:2013tjd}). We restrict ourselves to the region below the line, where $N^{\rm bkg} = 0.64$ electron recoil background events are expected while the neutron background is negligible. Of the $160$ observed events only one is found below the nuclear recoil mean.

The DM recoil rate below the nuclear recoil mean in the \LUX\ detector, assumed to be half of the total rate as discussed above, is computed as
\beq
R = \frac{1}{2} \int_{E^{\rm min}}^{E^{\rm max}} \hspace{-.33cm} \ud \ER \, \epsilon(\ER) \sum_T \frac{\ud R_T}{\ud \ER} \ ,
\eeq
with $E^{\rm min} = 3$ keV$_\text{nr}$ and $E^{\rm max} = 18$ keV$_\text{nr}$ the average lower and upper nuclear recoil energy threshold as quoted on page 41 of \Ref{GaitskellTalk}. We obtain the efficiency $\epsilon(\ER)$ by interpolating the black crosses in Fig.~9 of \Ref{Akerib:2013tjd}. The integrated form factor is therefore given by
\beq
\tilde{\mathcal{F}}_{i, j}^{(N,N')}(m_\chi)\rfloor_{\text{\LUX}} = \frac{1}{2} w \sum_T \xi_T
\int_{E^{\rm min}}^{E^{\rm max}} \hspace{-.33cm} \ud \ER \, \epsilon(\ER) \, \mathcal{F}_{i, j}^{(N, N')}(\ER, T) \ .
\eeq

\begin{figure}[!b]
\includegraphics[width=0.282\textwidth]{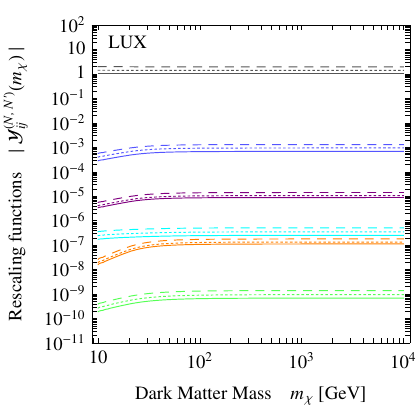} \quad
\includegraphics[width=0.182\textwidth, trim=0.7cm 0cm 2.1cm 0cm, clip=true]{Plots/legenda_Yieqj1}
\includegraphics[width=0.282\textwidth]{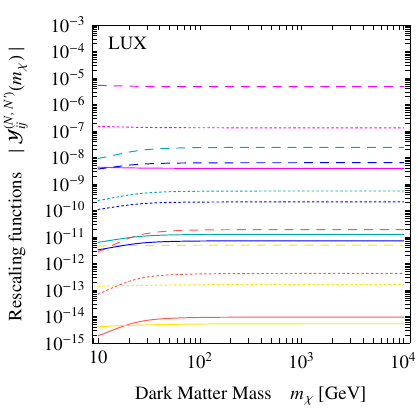} \quad
\includegraphics[width=0.182\textwidth, trim=0.7cm 0cm 2.1cm 0cm, clip=true]{Plots/legenda_Yieqj2} \\
\includegraphics[width=0.282\textwidth]{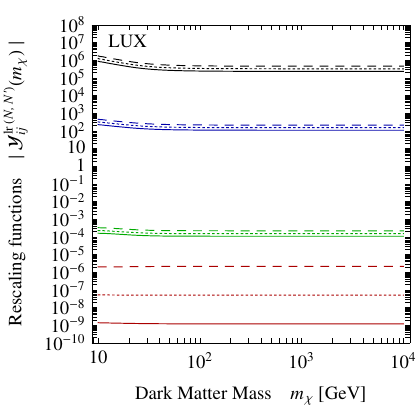} \quad
\includegraphics[width=0.182\textwidth, trim=0.7cm 0cm 2.1cm 0cm, clip=true]{Plots/legenda_Yieqj_LR}
\includegraphics[width=0.282\textwidth]{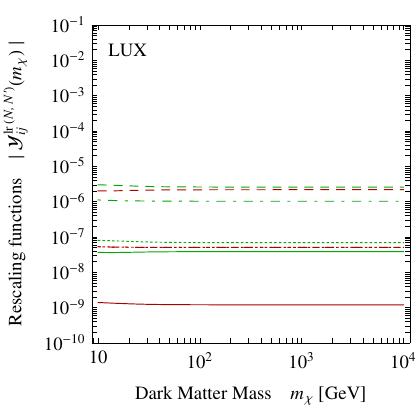} \quad
\includegraphics[width=0.182\textwidth, trim=0.7cm 0cm 2.1cm 0cm, clip=true]{Plots/legenda_Yineqj_LR} \\
\centerline{
\includegraphics[width=0.282\textwidth]{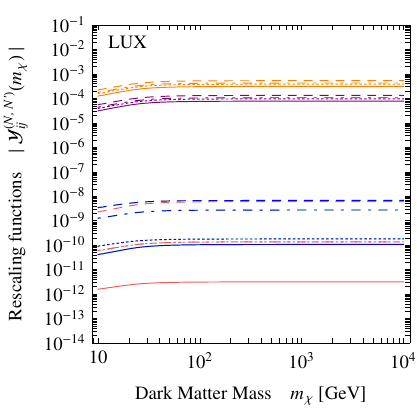} \quad
\includegraphics[width=0.182\textwidth, trim=0.7cm 0cm 2.1cm 0cm, clip=true]{Plots/legenda_Yineqj}
}
\caption{\em \small \label{fig:LUX} 
Absolute value of the {\bfseries rescaling functions} $\mathcal{Y}_{i, j}^{(N,N')}(m_\chi)$ and $\mathcal{Y}_{i, j}^{{\rm lr}(N,N')}(m_\chi)$ for the \LUX\ experiment.}
\end{figure}

The rescaling functions defined in \Eq{Y}, $\mathcal{Y}_{i, j}^{(N,N')}(m_\chi)$ for contact operators and $\mathcal{Y}_{i, j}^{{\rm lr}(N,N')}(m_\chi)$ for long range operators, are plotted for \LUX\ in \Fig{fig:LUX}.

\subsection*{Results and examples adding \LUX}
\addcontentsline{toc}{subsection}{Results and examples adding \LUX}

We have modified \Fig{fig:lambdaB}, \Fig{fig:boundsindividual}, \Fig{fig:standardbounds} and \Fig{fig:MDMinspired} in the main text to add the \LUX\ bound.

{\small 
\paragraph{Acknowledgments.}

\noindent E.D.N.~thanks Emilija Pantic for useful discussions.}

\newpage\normalsize

\section{Addendum: SuperCDMS 2014 data}
\label{SuperCDMS}

On February 28$^\text{th}$, 2014, the \SCDMS\ collaboration reported the result of a first search for DM particles using the \SCDMS\ detectors \cite{Agnese:2014aze}. An earlier analysis \cite{Agnese:2013jaa} had employed only one of the detectors, operated in a new mode ({\sc Cdmslite}, for {\sc Cdms} low ionization threshold experiment) that allowed to obtain an exquisitely low threshold thus setting a limit on DM-nucleon scattering cross sections for very light DM particles. In comparison, the present analysis, which employs data collected between October 2012 and June 2013, allows to set more stringent bounds in the high DM mass region. While the \LUX\ bound on heavy DM particles \cite{Akerib:2013tjd} is at present the strongest one for usual spin-independent scattering, the \SCDMS\ limit could become relevant when other interactions are taken into account, \eg isospin-violating DM \cite{Feng:2011vu}. Therefore, we include the \SCDMS\ result in our \href{http://www.marcocirelli.net/NRopsDD.html}{set of numerical tools} (Release 3) to derive bounds from direct DM searches.

\subsection*{Description of \SCDMS}
\addcontentsline{toc}{subsection}{Description of \SCDMS}

\SCDMS\ is an upgrade of the Cryogenic Dark Matter Search ({\sc Cdms II}) experiment (see Sec.~\ref{CDMS}), and is operated at the Soudan Underground Laboratory as its predecessor. The experiment consists of 15 germanium target crystals, each instrumented with ionization and phonon detectors. The measured ionization and phonon energies can be used to derive the recoil energy and the \virg{ionization yield}, \ie the ionization to recoil energy ratio which is used to distinguish signal from background.

A blind analysis with an exposure $w = 577$ kg$\,\cdot\,$days, using only the seven detectors with the lowest trigger thresholds, revealed 11 candidate events in the  range $1.6 - 10$ keV$_{\rm nr}$ \cite{Agnese:2014aze}. The background prediction is $N^{\rm bkg} = 6.1^{+1.1}_{-0.8}$ (stat + syst) events, with a negligible additional $0.098 \pm 0.015$ (stat) events from radiogenic and cosmogenic neutrons. The collaboration has however reason to believe that the adopted background model does not correctly account for a feature in one of the detectors, thus misestimating the background. The decision was made, prior to unblinding the data, to report an upper limit on the DM-nucleon scattering cross section.

The DM recoil rate is
\beq
R = \int_{E^{\rm min}}^{E^{\rm max}} \hspace{-.33cm} \ud \ER \, \epsilon(\ER) \sum_T \frac{\ud R_T}{\ud \ER} \ ,
\eeq
with $E^{\rm min} = 1.6$ keV$_\text{nr}$ and $E^{\rm max} = 10$ keV$_\text{nr}$, and the efficiency $\epsilon(\ER)$ taken to be the red line in Fig.~1 of \cite{Agnese:2014aze}. The integrated form factor is given by
\beq
\tilde{\mathcal{F}}_{i, j}^{(N,N')}(m_\chi)\rfloor_{\text{\SCDMS}} = w \sum_T \xi_T
\int_{E^{\rm min}}^{E^{\rm max}} \hspace{-.33cm} \ud \ER \, \epsilon(\ER) \, \mathcal{F}_{i, j}^{(N, N')}(\ER, T) \ .
\eeq

\begin{figure}[!b]
\includegraphics[width=0.282\textwidth]{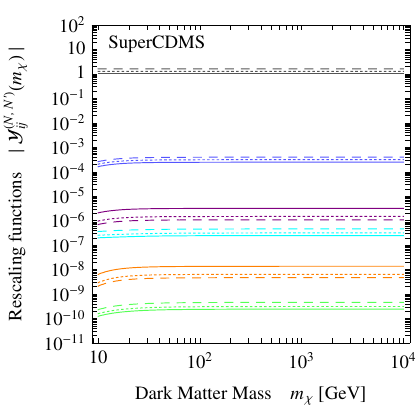} \quad
\includegraphics[width=0.182\textwidth, trim=0.7cm 0cm 2.1cm 0cm, clip=true]{Plots/legenda_Yieqj1}
\includegraphics[width=0.282\textwidth]{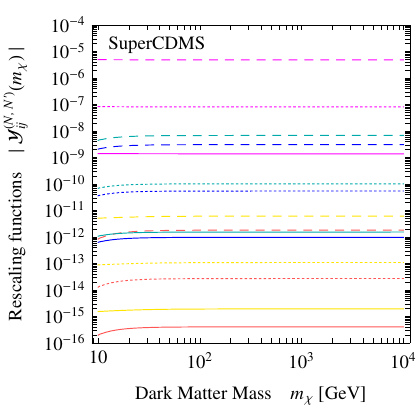} \quad
\includegraphics[width=0.182\textwidth, trim=0.7cm 0cm 2.1cm 0cm, clip=true]{Plots/legenda_Yieqj2} \\
\includegraphics[width=0.282\textwidth]{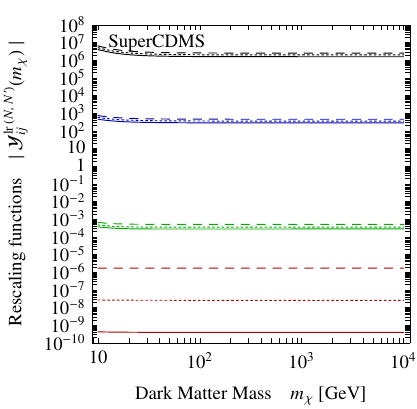} \quad
\includegraphics[width=0.182\textwidth, trim=0.7cm 0cm 2.1cm 0cm, clip=true]{Plots/legenda_Yieqj_LR}
\includegraphics[width=0.282\textwidth]{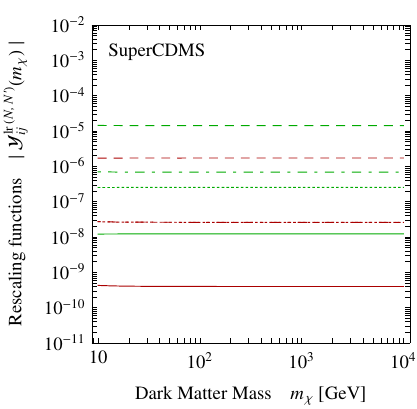} \quad
\includegraphics[width=0.182\textwidth, trim=0.7cm 0cm 2.1cm 0cm, clip=true]{Plots/legenda_Yineqj_LR} \\
\centerline{
\includegraphics[width=0.282\textwidth]{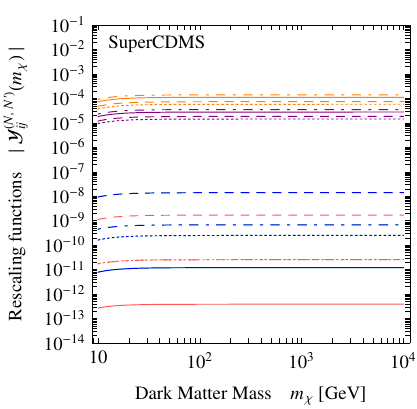} \quad
\includegraphics[width=0.182\textwidth, trim=0.7cm 0cm 2.1cm 0cm, clip=true]{Plots/legenda_Yineqj}
}
\caption{\em \small \label{fig:SCDMS} 
Absolute value of the {\bfseries rescaling functions} $\mathcal{Y}_{i, j}^{(N,N')}(m_\chi)$ and $\mathcal{Y}_{i, j}^{{\rm lr}(N,N')}(m_\chi)$ for the \SCDMS\ experiment.}
\end{figure}

The rescaling functions defined in \Eq{Y}, $\mathcal{Y}_{i, j}^{(N,N')}(m_\chi)$ for contact operators and $\mathcal{Y}_{i, j}^{{\rm lr}(N,N')}(m_\chi)$ for long range operators, are plotted for \SCDMS\ in \Fig{fig:SCDMS}.

\subsection*{Results and examples adding \SCDMS}
\addcontentsline{toc}{subsection}{Results and examples adding \SCDMS}

We have modified \Fig{fig:lambdaB}, \Fig{fig:boundsindividual}, \Fig{fig:standardbounds} and \Fig{fig:MDMinspired} in the main text to add the \SCDMS\ bound.

\end{document}